\documentclass[fleqn,usenatbib]{mnras}

\usepackage{newtxtext,newtxmath}
\usepackage[T1]{fontenc}
\usepackage[dvipsnames,usenames]{xcolor}
\DeclareRobustCommand{\VAN}[3]{#2}
\let\VANthebibliography\thebibliography
\def\thebibliography{\DeclareRobustCommand{\VAN}[3]{##3}\VANthebibliography}

\usepackage{graphicx}
\usepackage{amsmath}	
\usepackage{natbib}
\bibpunct{(}{)}{;}{a}{}{,}
\usepackage{colortbl}
\usepackage{xspace}
\usepackage{multirow,array}
\usepackage{anyfontsize}
\usepackage{comment}

\newcommand{\ic}{IceCube--200615A\xspace}

\newcommand{\Swift}{\textsl{Swift}\xspace}

\newcommand{\Fermi}{\textsl{Fermi}\xspace}

\newcommand{\nus}{\textsl{NuSTAR}\xspace}

\newcommand{\rxs}{1RXS J093117.6+033146\xspace}
\graphicspath{{./}{figures/}}
\title[Candidate counterpart to \ic]{A Radio-quiet AGN as a candidate counterpart to neutrino event \ic}
\author[F. McBride et al.]{Felicia McBride$^{1}$,\thanks{f.mcbride@bowdoin.edu}
Nur Schettino$^{1,2}$,
John D. O'Brien$^{3}$,
Wilder Harwood$^{1,4}$,
Liz Perot$^{1}$,
Grant Temple$^{5}$,
\newauthor
Hugo Ayalo Solares$^{6}$,
Alessandra Corsi$^{7,8}$,
Alexis Coleiro$^{9}$,
Doug Cowen$^{6}$,
Derek B. Fox$^{5}$,
Yijia Li$^{5}$,
\newauthor
Kohta Murase$^{6}$,
Andrew Pellegrino$^{5}$,
Thomas D. Russell$^{10}$
Stephanie Wissel$^{6}$
\\
\\
$^{1}$ Department of Physics and Astronomy, Bowdoin College, College Sta 8800, Brunswick, Maine 04011, USA\\
$^{2}$ Dr. Karl Remeis Sternwarte/ECAP at Friedich-Alexander Universit\"at Erlangen-N\"urnberg, Sternwartstr. 7, Bamberg, Germany\\
$^{3}$ Department of Mathematics, Bowdoin College, 8600 College Sta, Brunswick, Maine 04011, USA\\
$^{4}$ Department of Physics, Brown University, Providence, RI 02912, USA\\
$^{5}$ Department of Astronomy \& Astrophysics, The Pennsylvania State University, University Park, PA 16802, USA\\
$^{6}$ Department of Physics, The Pennsylvania State University, University Park, PA 16802, USA\\
$^{7}$ Department of Physics and Astronomy, Texas Tech University, Box 1051, Lubbock, TX 79409-1051, USA\\
$^{8}$ William H. Miller III Department of Physics and Astronomy, Johns Hopkins University, Baltimore, MD 21218, USA\\
$^{9}$ APC/Universit\'e Paris Cit\'e, CNRS, Astroparticule et Cosmologie, F-75013 Paris, France\\
$^{10}$ INAF, Istituto di Astrofisica Spaziale e Fisica Cosmica, Via U. La Malfa 153, I-90146 Palermo, Italy
} 
\date{Accepted 24-06-2025. Received YYY; in original form ZZZ}
\pubyear{2025}
\begin{document}
\label{firstpage}
\pagerange{\pageref{firstpage}--\pageref{lastpage}}
\maketitle
\begin{abstract}
Follow-up observations of neutrino events have been a promising method for identifying sources of very-high-energy cosmic rays. Neutrinos are unambiguous tracers of hadronic interactions and cosmic rays.
On June 15, 2020, IceCube detected a neutrino event with an 82.8\% probability of being astrophysical in origin.
 To identify the astrophysical source of the neutrino, we used X-ray tiling observations to identify potential counterpart sources. We performed additional multiwavelength follow-up with \nus and the VLA in order to construct a broadband spectral energy distribution (SED) of the most likely counterpart. 
From the SED, we calculate an estimate for the neutrinos we expect to detect from the source. 
While the source does not have a high predicted neutrino flux, it is still a plausible neutrino emitter. It is important to note that the other bright X-ray candidate sources consistent with the neutrino event are also radio-quiet AGN.
A statistical analysis shows that \rxs is the most likely counterpart (87.5\%) if the neutrino is cosmic in origin and if it is among X-ray detectable sources. This results adds to previous results suggesting a connection between radio-quiet AGN and IceCube neutrino events.
\end{abstract}



\section{Introduction}
The sources of very-high-energy and ultra-high-energy cosmic rays have not been conclusively identified \citep[e.g.,][]{Norman1995}, although recent progress has been made with the identification of possible origins of neutrino events \citep{Kadler2016,txs0506,tde,ngc1068,ic1903}.
Cosmic rays consist of protons and nuclei and are observed at energies of $10^{9}$~--~$\sim10^{20}$\,eV.
It is not fully clear what processes are able to accelerate cosmic rays to extreme energies; proposed mechanisms include astrophysical shocks \citep{Bell1978} and magnetic reconnection \citep{diMatteo1998}.
The origin of cosmic rays cannot be traced back to their sources directly as they
are deflected by magnetic fields, including those located near their origin, the Galactic magnetic field and Earth's magnetic field.

If protons interact with protons or photons close to their source of origin, secondary particles (such as neutrinos) could allow us to identify cosmic-ray sources \citep{Halzen2002}. Proton-photon interactions produce neutral and charged pions.
Neutral pions decay into photons, which we expect to see in the X-ray
and $\gamma$-ray bands, while charged pions decay into a subsequent
cascade of particles that include neutrinos \citep[e.g., ][]{Mannheim1995}. 
As neutral particles, we expect neutrinos produced at their source of origin to be tracers of cosmic-ray sources. 
Dense regions near the production sites of neutrinos may be opaque to high and very-high energy $\gamma$ rays, i.e., $\gamma$-ray emission is not necessarily a good predictor for expected neutrino flux \citep{Murase2016,Gao2017,Krauss2018,Gao2019}.\\
Neutrinos can also be created in Earth's atmosphere, due to the interaction of cosmic rays with nuclei in the atmosphere. As the incoming cosmic rays can be deflected by galactic and intergalactic magnetic fields, atmospheric neutrinos do not reveal their cosmic origin.
Atmospheric neutrinos and the positional reconstruction of neutrino events both pose separate challenges to a neutrino's association with an astrophysical source. For the latter, reconstructing a neutrino's path in the IceCube detector can produce large angular uncertainties depending on neutral vs charged current interactions and neutrino flavour in the detector volume \citep{IC2017}.

Several different types of sources have been proposed as potential sites of proton acceleration and neutrino production at very-high and ultra-high energies, including Gamma-ray bursts \citep[GRBs;][]{Waxman1997,Vietri1998} that may have choked jets \citep{Murase2013b}, Tidal Disruption Events \citep[TDEs;][]{Farrar2009,Dai2017}, and star-forming/starburst galaxies \citep{Loeb2006,Murase2013}. Neutrino production has also been 
predicted in the cores \citep{Eichler1979,Berezinskii1981, Protheroe1983,Begelman1990,Stecker1991} and the jets of active galactic nuclei
\citep[AGN;][]{Biermann1987,Mannheim1992,Mannheim1993,Rachen1993}.

IceCube is a neutrino detector at the South Pole that has been operating since 2010 \citep{ic1}. It has confirmed the detection of a flux of astrophysical neutrinos \citep{IC2013,Aartsen2014}. 
IceCube detects two types of events, cascade and track events. Cascade events originate from neutral current interactions as well as charged current interactions (except for $\nu_\mu$), the energy is deposited within the detector volume within a small roughly spherical area, while charged current $\nu_\mu$ interactions leave a track \citep{Chirkin2004,IC-angrecon}.
Cascade events have large uncertainty regions, it is difficult to identify possible sources \citep{Krauss2014, Krauss2018}.
The first promising associations between a neutrino event and a blazar was
suggested by \citet{Halzen2005}, connecting a flare of the blazar 1ES\,1959+650 with AMANDA data.
A similar association was proposed in 2016, with the flaring blazar PKS\,2155$-$304 and the ``Big Bird'' neutrino event IceCube-35 \citep{Kadler2016}. 
This was an event at 2.0\,PeV with a high ``signalness'' (likelihood that the neutrino is astrophysical in origin). Unfortunately, the angular uncertainty of the event is large with a radius of $15\fdg9$ at a 50\% confidence level.
This was followed by another blazar-neutrino association, IceCube-170922A and TXS\,0506+056. 
IceCube-170922A was a track event at a much smaller angular
uncertainty, but at much lower signalness of 
$56.5$\%\footnote{\url{https://gcn.gsfc.nasa.gov/notices_amon/50579430_130033.amon}} \citep{txs0506}. TXS 0506+056 exhibited a brightening (flare) in
the $\gamma$-ray band during the neutrino event; this allowed the two events to be connected statistically at a 3$\sigma$ confidence level. This value was determined by investigating how many blazars in the \Fermi/LAT band are flaring at any given point. The confidence level of the association does not consider the signalness of the event. 
A previous low-energy cluster of neutrinos was detected from the position of TXS\,0506+056
in March 2015, but no change in $\gamma$-ray flux was detected by \Fermi/LAT \citep{IC2018}.
In TXS\,0506+056, X-ray observations in the soft and hard X-ray band
were crucial in constraining the hadronic contribution in order to
determine whether the source is a likely neutrino counterpart
\citep{Keivani2018,Gao2019}.
The results from TXS\,0506+056 suggested that most of the X-ray and
$\gamma$-ray radiation must have a leptonic origin. 

In AGN jets, the low-energy peak is created by $e^\pm$ synchrotron radiation,
while the high-energy peak can be explained by Inverse Compton upscattering by relativistic electrons \citep{Ginzburg1965,Rees1967, Dermer1993, Sikora1994}.
The photon seed field for the upscattering can either be the same population of Synchrotron photons \citep[Synchrotron Self Compton; SSC; e.g.][]{Maraschi1992} or other photon fields, including thermal emission from e.g., the accretion disk, the Broad Line Region (BLR) or the torus \citep[External Compton; EC;][]{Tavecchio2000}.
It is expected that both leptonic and hadronic processes contribute to
the high-energy SED. Spectral modelling alone is not able to distinguish
between leptonic and hadronic models conclusively \citep{Boettcher2013}.

\citet{Krauss2018} investigated the ``hadronicness'', i.e., the
expected contribution to blazar high-energy peaks in the SED, which indicates that low hadronic contributions of 7.9\% are expected.
That is, less than 8\% of the high-energy emission of blazar SEDs can originate from hadronic processes, otherwise much higher numbers of neutrinos should be detected by IceCube. 
This average assumes that this number is the same for all blazars, which may not be the case. There may be differences in flat-spectrum radio quasars (FSRQs) vs. BL Lac sources or in individual sources; i.e. one blazar may have a much higher ``hadronicness'' than another source and therefore produce a much higher number of neutrinos. This also assumes that the neutrino spectrum produced by blazars is covered entirely by the IceCube detector. This is unlikely, and we expect significant contributions to the neutrino spectrum at energies of the order of $10^{18}$\,eV \citep{Mannheim1995}.
Electrons and positrons are lighter and therefore more likely to be
accelerated to (ultra-)relativistic energies. Leptonic emission should contribute strongly to the high-energy SEDs. As the origin of the high-energy emission cannot be disentangled, this complicates the calculation of an expected neutrino flux.

Following the identification of possible blazars with neutrinos, Seyferts (radio-quiet AGN that may or may not be jetted) were identified as a potential source: the neutrino event IceCube-190331A was found to be positionally consistent with a Seyfert type 1.2 AGN \citep{Krauss2020}.
In this case, neutrinos are expected to be created in or near the corona of AGN \citep{Stecker1991,Murase2020,Kheirandish2021} or in a weak jet \citep{AlvarezMuniz2004}.
The recently discovered neutrino source NGC\,1068 \cite{ngc1068} is a Seyfert type 2 AGN with a weak jet \citep{Cotton2008}.
Another new result is a contribution to the IceCube flux from Galactic sources \citep{ICgal}.

In this work, we have investigated the IceCube neutrino event IceCube-200615A. It was detected with a signalness of 82.83\% and a positional uncertainty (at a 90\% confidence level) of 55.2\,$^{\prime}$.
We performed follow-up observations to identify possible counterparts.
Due to the COVID-19 pandemic, many observatories and facilities were closed; this limited the multiwavelength follow-up drastically, preventing us from obtaining optical follow-up to further identify the counterpart.

In this paper, we present potential astrophysical source counterparts and identify the most likely counterpart. In Section \ref{sec:obs}, we present the follow-up observations. Section \ref{sec:methods} describes the methods used to
identify a likely neutrino counterpart and compile a broadband SED.
In Section \ref{sec:results}, we discuss the results, including the calculation of the neutrino flux of the likely counterpart.
The neutrino uncertainty area is discussed in Section \ref{sec:disc}, while Section \ref{sec:stat} highlights a statistical analysis of the source counterparts. Section~\ref{sec:concl} concludes the main findings.
Throughout this paper we use a $\Lambda$CDM cosmology with $\Omega_{\mathrm{m}}=0.3$, $\Lambda=0.7$,
$H_0=70$\,km\,s$^{-1}$\,Mpc$^{-1}$ \citep{Beringer2012}.
All coordinates are given with a J2000.0 equinox.

\section{Observations}
\label{sec:obs}
In this Section, we present information regarding the neutrino event IceCube-200615A, and the subsequent multiwavelength follow-up observations, including \Swift and \nus observations.

\subsection{IceCube Detection}
\label{subsec:obs-ic}
On June 15th 2020 at 14:49:17.38 UT, the IceCube Neutrino Observatory
located at the geographic South Pole, detected a high-energy neutrino
that originated from the coordinates right ascension $\alpha$ =
$142\fdg 95^{+1.18}_{-1.45}$ and declination $\delta$ = $3\fdg66^{+1.19}_{-1.06}$ (J2000; 90\% PSF containment). 
The radii of the positional uncertainty of the neutrino
are 21.5$^\prime$ (50\%) and 55.2$^\prime$ (90\%). This corresponds to
an area of approximately $0.403$\,deg$^2$ and
$2.659$\,deg$^2$, respectively \citep{IceCubeAlert}. This neutrino had a high probability of having an astrophysical origin with a signalness of 
82.832\% despite not being a HESE event\footnote{\url{https://gcn.gsfc.nasa.gov/notices\_amon\_g\_b/134191\_17593623.amon}}.
We selected this event for follow-up observations due to its high signalness and small angular uncertainty.

\subsection{Swift/XRT}
\label{subsec:obs-sw}
Following the detection of the neutrino event, we requested follow-up observations with the \textsl{Neil Gehrels Swift Observatory} \citep{swift} within the Astrophysical Multimessenger Observatory Network
(AMON)\footnote{\url{https://www.amon.psu.edu/}} framework \citep{amon}.
Observations began on June~15, 18:59:16~UT (4 hours 9 minutes after the initial neutrino detection) and ended June~16 11:08:53~UT (20 hours 19 minutes after the neutrino detection). 
A 7-point tiling pattern covered about 1\,deg$^2$, 37.6\% of the 90\%
uncertainty region, but more than twice the area of the 50\% uncertainty
region. 

We applied the newest calibration to the data using \texttt{xrtpipeline} \citep[HEASoft V.6.29;][]{heasoft}.
Products were generated with \texttt{ximage}, which also allowed us to identify all X-ray source with a SNR$\ge\,3$. 
Nine potential X-ray sources were detected by \citep{AMONAlert}. However, the low signal-to-noise ratio led to some spurious detections, which are unlikely to be astrophysical sources. See Section \ref{subsec:srcs} and Table~\ref{tab:srcs} for a detailed discussions of all X-ray sources.

\rxs is the brightest X-ray source out of four significant detections in the X-ray telescope (XRT) observations and therefore a likely counterpart. X-ray brightness is related to neutrino brightness due to expected pion cascade production of X-ray emission; see Section~\ref{subsec:scale} for details.
Data for \rxs was extracted with a source region centred on coordinates $\alpha=431\fdg7603$, $\delta=744\fdg6023$ with a radius of $54.218^{\prime\prime}$, while the background region was an annulus centred on the same coordinates with inner and outer radii $82.506^{\prime\prime}$ and
$235.731^{\prime\prime}$.

\subsection{NuSTAR}
\label{subsec:obs-nu}
Following the identification of \rxs as the brightest source in the \Swift/XRT observations and therefore a potential counterpart, we
triggered our \nus program. 
The \textsl{Nuclear Spectroscopic Telescope Array} (\nus) detects photons in the 3--79\,keV band \citep{nustar}, which allows us to constrain the hadronic contribution to the broadband SED.

\nus observed \rxs on 2020 June 16 at 22:31:09 UT (1 day 8 hours and 41 minutes after the neutrino detection).
The data obtained for \rxs were extracted using \texttt{nupipeline} v0.4.8. The source region, obtained from ds9 was selected with coordinates $\alpha=142\fdg8270833$ and $\delta = 3\fdg5197222$ for focal plane A.
The focal plane B source region was selected for coordinates centred on $\alpha = 142\fdg825$ and $\delta = 3\fdg5208333$.
Both were extracted with a radius of $50^{\prime\prime}$.
The background regions were selected, centred on coordinates $\alpha= 142\fdg7541667$ and $\delta= 3\fdg4527778$ for focal plane A.
Focal plane B background region was centred on source coordinates
$\alpha = 142\fdg7416667$ and $\delta=3\fdg4638889$.
The extraction radius for the background was 215$^{\prime\prime}$.
After creating source and background regions for both observations, \texttt{nuproducts} v0.3.2 was used to extract the spectrum pha files.

\subsection{Radio/VLA}
\label{subsec:obs-vla}
We performed radio follow-up with the Karl G. Jansky Very Large Array \citep[VLA;][]{vla}, under project code 20A-586 (PI: F. McBride). 
Observations of the target field were taken on 2020-08-25 between 18:05:45 and 18:43:18 UT (71 days after the initial neutrino detection),
during which the array was in its B-configuration%
\footnote{\url{https://public.nrao.edu/vla-configurations/}}. 
Data were recorded in 8-bit mode at C-band, with two 1-GHz base-bands centred at 4.5 and 7.5\,GHz. Bandpass and flux calibration was done with 3C\,147 (J0542$+$4951), while J0925$+$0019 was used for phase calibration. 
Calibration and imaging was carried out following standard procedures with the Common Astronomy Software Package (\textsc{casa}, version 5.1.3; \citealt{casanew}). To maximise sensitivity, we used natural weighting when imaging, resulting in angular resolutions of 1.84$^{\prime\prime}\times$1.46$^{\prime\prime}$ with a position angle of 19$^\circ$ East of North at 4.5 GHz and 1.02$^{\prime\prime}\times$0.79$^{\prime\prime}$ (position angle 18$^\circ$ East of North) at 7.5 GHz.

\begin{figure}
    \centering
    \includegraphics[width=0.5\textwidth]{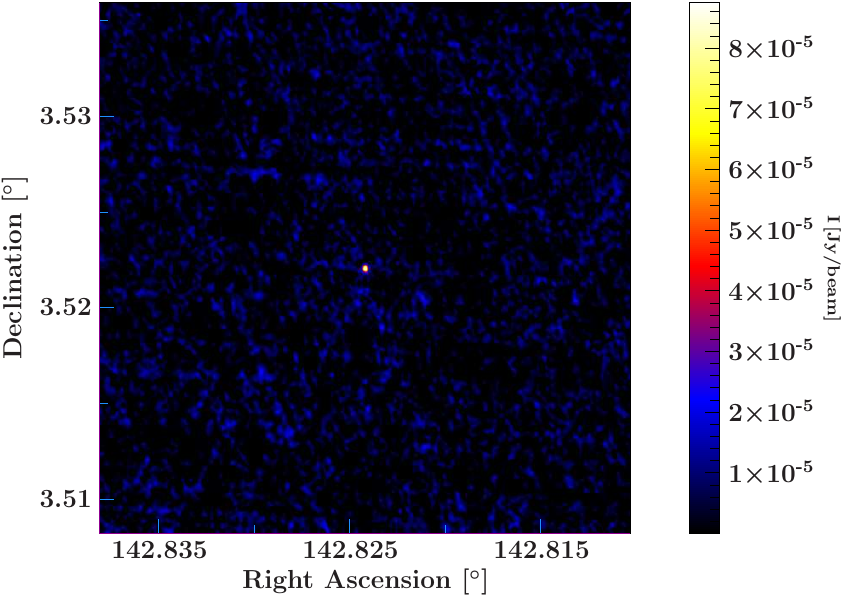}
    \caption{VLA C-band image of \rxs.}
    \label{fig:vla}
\end{figure}

The radio counterpart to \rxs was detected at both central frequencies. Fitting for a point source in the image plane, we measured radio flux densities of $94 \pm 8$ $\mu$Jy at 4.5 GHz and $88 \pm 6$ $\mu$Jy at 7.5 GHz. We determine the radio spectral index, $\alpha$, to be $-0.13 \pm 0.25$, where the radio flux density $S_{\nu}$ is proportional to the observing frequency $\nu$ such that $S_{\nu} \propto \nu^{\alpha}$.
The VLA detection is presented in Fig.~\ref{fig:vla}.

\subsection{\Swift/UVOT}
\Swift performed observations using its UltraViolet-Optical Telescope
(UVOT) simultaneous to the X-ray observations \citep{uvot}. Observations were performed with 7 tiles in the U filter with
an average exposure time of 348\,s, UVOT was able to detect 5
sources (1, 3, 4, 6, and 9) found by XRT as listed by \citet{AMONAlert}.  
This includes source \#1 which is \rxs. Source 2 was not
detected by UVOT suggesting that the XRT detection was spurious and sources 5, 7, and 8 all were outside the the
UVOT field of view \citep{UVOTAlert}.
Data for \rxs was extracted using a 5$^{\prime\prime}$ region
centred on coordinates $\alpha=142\fdg8242$, $\delta=3\fdg5221$, with an annulus background
region centred on the same coordinates with inner and outer radii of $13^{\prime\prime}$ and $26^{\prime\prime}$.

\subsection{\Fermi/LAT}
The Large Area Telescope onboard the \textsl{Fermi Gamma-ray Space Telescope} continuously monitors the sky \citep{fermilat} and is therefore able to provide near simultaneous data.
\Fermi/LAT data shows that no Gamma-ray Bursts were detected near the neutrino event time. Additionally,
\Fermi/LAT detected no source in the neutrino uncertainty region. This results in a flux upper limit of $3\times 10^{-10}$\,ph\,cm$^{-2}$\,s$^{-1}$ \citep[95\% confidence;][]{FermiAlert}. We use this upper limit to constrain our broadband spectral model in the energy range of 100\,MeV to 300\,GeV (see Sect.~\ref{subsec:sed}).

\subsection{HAWC}
The High-Altitude Water Cherenkov array \citep{hawc} was unable to detect \rxs
at very-high energies in the 300\,GeV -- 100\,TeV band. The upper
limit of the HAWC observations yield an upper limit of $3.08\times 10^{-13}
\frac{E}{\mathrm{TeV}}^{-0.3}$\, TeV\, cm$^{-2}$\, s$^{-1}$
\citep{HAWCAlert}, which we use to constrain our SED model.

\subsection{Archival Data}
After identifying a possible counterpart to IceCube-200615A, archival
observational data were used when analysing the multiwavelength
emissions from \rxs.
2MASS\,J09311779+0331194 is the most likely optical counterpart to \rxs at an angular distance of 0.084$^{\prime\prime}$. Archival data for this source includes data from the following catalogues: GALEX \citep{galex}, ALLWISE \citep{wise}, US Naval Observatory A2 Catalog (USNO A2.0)\footnote{http://tdc-www.harvard.edu/catalogs/usnosa2.html}, Sloan Digital Sky Survey (SDSS), 2MASS \citep{2MASS}.

Many of these observations were taken months or years before the neutrino event and are not considered (quasi-)simultaneous to the SED data. 
We exclude these data from the model and only add it for display purposes 
and to investigate variability. Archival data is presented in black colour in Fig.~\ref{fig:sed}.

\section{Methods}
\label{sec:methods}

\subsection{Broadband Spectral Energy Distribution (SED)}
\label{subsec:sed}
From the multiwavelength data described in Section~\ref{sec:obs}, we compiled a multiwavelength broadband SED. We used the
\textsl{Interactive Spectral Interpretation System (ISIS)} to analyse
the multiwavelength data. X-ray and \Swift/UVOT data are treated in
detector space. Data obtained in flux units were assigned a diagonal
response. ISIS allows us to use a model with a forward folding approach \citep{ISIS}.
Unfolding of the data for displaying purposes in the SED uses a model-independent flux calculation.
For the photoelectric absorption in the X-ray band we used \texttt{tbabs} with the \texttt{vern} cross-sections \citep{Verner1996}
and the \texttt{wilm} abundances \citep{Wilms2000}. 
We froze the neutral hydrogen equivalent absorbing column $N_{\mathrm{H}}$ to the Galactic value obtained from the HI4PI survey \citep{HI4PI}, $N_{\mathrm{H}}= 2.78\times 10^{20}\,$cm$^{-2}$. 
No evidence for additional absorption was present in the X-ray data.
Infrared, optical, and ultraviolet data was corrected for extinction 
using the same neutral hydrogen-equivalent absorbing column 
$N_{\mathrm{H}}$ used for \texttt{tbabs}. This method follows
\citet{Nowak2012}.
The SED was modelled with an empirical logarithmic parabola
\citep{Massaro2004}.
$\Phi_{\gamma}$, the integrate energy flux was obtained from
integrating the high-energy peak of the model in the 0.1\,keV -- 300\,GeV energy band.
For details of the SED generation and analysis see Sections~2.4 and 2.5 of \citet{Krauss2016}.
It is worth noting that we apply $\chi^2$ statistics to the model. Due to the lack of cross-calibration between instruments it is unclear how reliable this method is. We use the $\chi^2$ statistics not for the absolute goodness-of-fit, but for a relative comparison between model parameters. 

\subsection{Energies}
In this section we discuss the energies of the photons and the neutrinos. Pion-photoproduction occurs when a UV photon interacts with the protons accelerated near a supermassive black hole.
UV photons could originate from an accretion disk. In the comoving frame of the proton, the UV photons would be redshifted at $E^\prime=E/\Gamma$ (from the inner accretion disk) or blueshifted (scattered photons). 
Neutrinos carry away 5--10\% of the proton energy, which results in $E_\nu\sim0.1\Gamma (E^\prime/(30\,\mathrm{eV}))^{-1}\,\mathrm{PeV}$ 
in the observer's frame. For expected values of $E=30\,$eV and $\Gamma=10$, the neutrino range covers the energy range of 100\,TeV--10\,PeV. 
The exact energies are uncertain, due to the uncertainty in the Lorentz factor of accelerated protons in the vicinity of supermassive black holes.

\subsection{Expected neutrino numbers}
\label{subsec:nunum}
We based our approach on Sections 2.5 and 2.6 by \citet{Krauss2018}, which describe the calculation
of an expected neutrino flux. Eq.~4 parametrizes the neutrino
spectrum as $F_\nu(E_\nu) = C\cdot E^{1-\Gamma_{\nu}}$.
While this is a
correct mathematical description, the correlation between integrated
electromagnetic (EM) energy flux and integrated neutrino energy flux requires the
definition of the energy flux ($E^2\cdot\frac{\mathrm{d}N}{\mathrm{d}E}$ instead of
$E\cdot \frac{\mathrm{d}N}{\mathrm{d}E}$).
The calculation is based on $$\Phi_\gamma = \Phi_\nu\qquad,$$
where $\Phi$ is the integrated photon or neutrino energy flux
\begin{equation}
\Phi_\gamma = \Phi_\nu =
\int_{E_{\nu,\mathrm{min}}}^{E_{\nu,\mathrm{max}}} E\cdot F_{\nu}
\mathrm{d}E\qquad.
\label{eq:def}
 \end{equation}
This assumption of equivalence of EM and neutrino fluxes is based on Monte Carlo (MC) simulations of pionphotoproduction performed by \citet{Muecke2000}.
The neutrino spectrum is currently uncertain for all sources. Due to
this lack of information, we assume a simple powerlaw shape for the
neutrino spectrum
\begin{equation}
E\cdot F_\nu (E_\nu) = C\cdot E_\nu^{2-\Gamma_\nu}\qquad.
\end{equation}
Integrating this Eq.~\ref{eq:def} yields
\begin{equation}
\Phi_\nu =\int_{E_{\nu,\mathrm{min}}}^{E_{\nu,\mathrm{max}}} C\cdot
E_\nu^{2-\Gamma_\nu} = \dfrac{1}{3-\Gamma_\nu}C\cdot E^{3-\Gamma_\nu}
\equiv \Phi_\gamma
\end{equation}
This allows us to solve for C:
\begin{equation}
  C = \dfrac{\Phi_{\gamma}~(3-\Gamma_{\nu})}{(E_{\mathrm{{max}}}^{3-\Gamma_{\nu}}-E^{3-\Gamma_{\nu}}_{\mathrm{min}})}\qquad,
  \label{eq:c}
\end{equation}
where $E_{\mathrm{min}}$ and $E_{\mathrm{max}}$ refer to the minimum and
maximum neutrino energy; i.e., the energy range in which the source emits
neutrinos. 

The (differential) number of observed neutrinos is then given by
\begin{equation}
\dfrac{\mathrm{d}N_{\nu}}{\mathrm{d}E_{\nu}} = \dfrac{F_{\nu}
  (E_{\nu})}{E_{\nu}}\cdot A_{\mathrm{eff}}(E_{\nu})\cdot T\qquad,
\end{equation}
with the effective area of IceCube $A_{\mathrm{eff}}$ at a given energy
$E_\nu$, the exposure time $T$ and the neutrino flux $F_\nu$. Using
the powerlaw equation and Eq.~\ref{eq:c} yields
\begin{equation}
E_\nu\cdot \dfrac{\mathrm{d}N_{\nu}}{\mathrm{d}E_{\nu}} =
\dfrac{\Phi_{\gamma}\cdot(3-\Gamma_{\nu})}{(E_{\mathrm{{max}}}^{3-\Gamma_{\nu}}-E^{3-\Gamma_{\nu}}_{\mathrm{min}})}\cdot
E^{1-\Gamma_{\nu}}\cdot T\cdot A_{\mathrm{eff}}(E_{\nu})\qquad.
\end{equation}

The effective area of IceCube is provided as a binned table of
energies and corresponding areas\footnote{\url{https://icecube.wisc.edu/data-releases/2020/02/all-sky-point-source-icecube-data-years-2012-2015/}}.

To calculate the total number of neutrinos, we integrate the neutrino
spectrum in each energy bin. The neutrino number in a single bin is defined
as
\begin{equation}\label{eq:int_bin}
\begin{split}
N_\nu (E_{\nu,i}) &= \int_{E_1}^{E_2} E_\nu \cdot
\dfrac{\mathrm{d}N}{\mathrm{d}E_\nu}  \mathrm{d}E_\nu \\[1em]
    &= \dfrac{C\cdot T}{2-\Gamma_{\nu}}\cdot
A_{\mathrm{eff}}(E_{\nu,i}) \cdot
(E^{2-\Gamma_{\nu}}_{\mathrm{max}}-E^{2-\Gamma_{\nu}}_{\mathrm{min}})\qquad ,
\end{split}
\end{equation}
where $E_{\nu,i}$ is the energy of one specific bin.

To calculate the total  neutrino number therefore requires summing
each energy bin.
\begin{equation}
\begin{split}
& \sum_{E_{\nu,i}} N_{\nu} (E_{\nu,i})\\[1em] &=
\sum_{E_{\nu,i}=E_{\mathrm{min}}}^{E_{\mathrm{max}}}  \dfrac{C\cdot T}{2-\Gamma_{\nu}}~~\cdot
A_{\mathrm{eff}}(E_{\nu,i}) \cdot\\[1em]
&\null (E^{2-\Gamma_{\nu}}_{\nu,i}-E^{2-\Gamma_{\nu}}_{\nu,i})
\end{split}
\label{eq:final}
\end{equation}

This approach is equivalent to the calculation by \citet{Krauss2018}, particularly Eq.~\ref{eq:final} in this manuscript can be compared directly to Eq.~8 in \citet{Krauss2018}.
The difference is only the input energy spectrum $E^{2-\Gamma_\nu}$ vs. $E^{1-\Gamma_\nu}$, which causes differences in the numbers calculated. We have tested both equations and found only small difference in the resulting numbers, suggesting the conclusions drawn by \citet{Krauss2018} are largely still valid.
For $\Gamma_\nu=1.9$ and $\beta_2 = 0.08$, Eq.~\ref{eq:final} results in $N_{\nu,\mathrm{max}}=0.408$. The Eq.~8 by \citet{Krauss2018} predicts $N_{\nu,\mathrm{max}}=0.472$. For $\Gamma_\nu=2.89$ and $\beta_2 = 0.08$, our approach here predicts $N_{\nu,\mathrm{max}}=0.471$, while the previous approach predicts $N_{\nu,\mathrm{max}}=0.461$. This shows differences in the predicted neutrino numbers of $\sim 2$ -- 13\%, which is negligible compared to the overall uncertainties.

\subsection{Neutrino scaling factors}
\label{subsec:scale}
The assumptions behind the neutrino number calculation in
Section~\ref{subsec:nunum} is that the total integrated
electromagnetic energy flux corresponds to the total integrated neutrino
energy flux. Only photons generated by the decay of neutral pions
are relevant for this calculation.
This excludes the synchrotron emission, i.e., the low-energy peak of
the SED. The high-energy peak is potentially created from the decay of
$\pi^{0}$. Due to this, $\Phi_{\gamma}$ is the integrated energy flux
of the EM spectrum in the X-ray and $\gamma$-ray band.

Two factors complicate the result:
\begin{enumerate}
\item We expect a strong leptonic contribution to the X-ray and
  $\gamma$-ray band. This means we are overpredicting the number of neutrinos by assuming the entire high-energy peak is hadronic.
\item The regions where protons and photons are able to interact to
  produce neutral pions may be opaque to high-energy photons, which
  could underpredict the number of neutrinos.
\end{enumerate}

Concerning (i): While stacking analyses have been able to place strong
limits on the contribution of neutrinos from blazars to the IceCube
data set \citep{stacking2017,stacking22}, there is no stringent limit for radio-quiet
AGN. Further, \citet{Krauss2018} compared the total number of neutrinos expected for
all $\gamma$-ray bright blazars to the number of neutrinos and found
that, for blazars, no more than $\sim$ 7.9\% of the
high-energy peak can originate from hadronic interactions;
otherwise, we would expect much higher numbers of neutrinos.
Such a limit does not exist for radio-quiet AGN (yet). Additionally, it is
unclear what fraction of the high-energy emission is expected to be
hadronic.
Instead of the blazar-specific scaling factors from \citet{Krauss2018}, we apply an optimistic scaling factor of 30\%, i.e., we calculate a scaled neutrino number $N_{\nu,\mathrm{scaled}}$. The expected scaling factor is larger in this case, as radio-quiet AGN tend to be weaker X-ray sources. An exact number is not available and will require a population study to constrain the hadronic contribution from radio-quiet AGN.
Concerning (ii): We expect that while the $\gamma$-ray flux may be
suppressed in regions where neutrinos are generated, photons might be reprocessed towards lower energies, which predicts higher flux in the X-ray or MeV band \citep[e.g.,][]{Murase2016}. This suggest that the best path forward is to identify the X-ray brightest sources and investigate them as potential neutrino counterpart sources.

\section{Results}
\label{sec:results}

\subsection{\Swift/XRT Sources}
\label{subsec:srcs}

\begin{figure}
    \centering
    \includegraphics[width=0.5\textwidth]{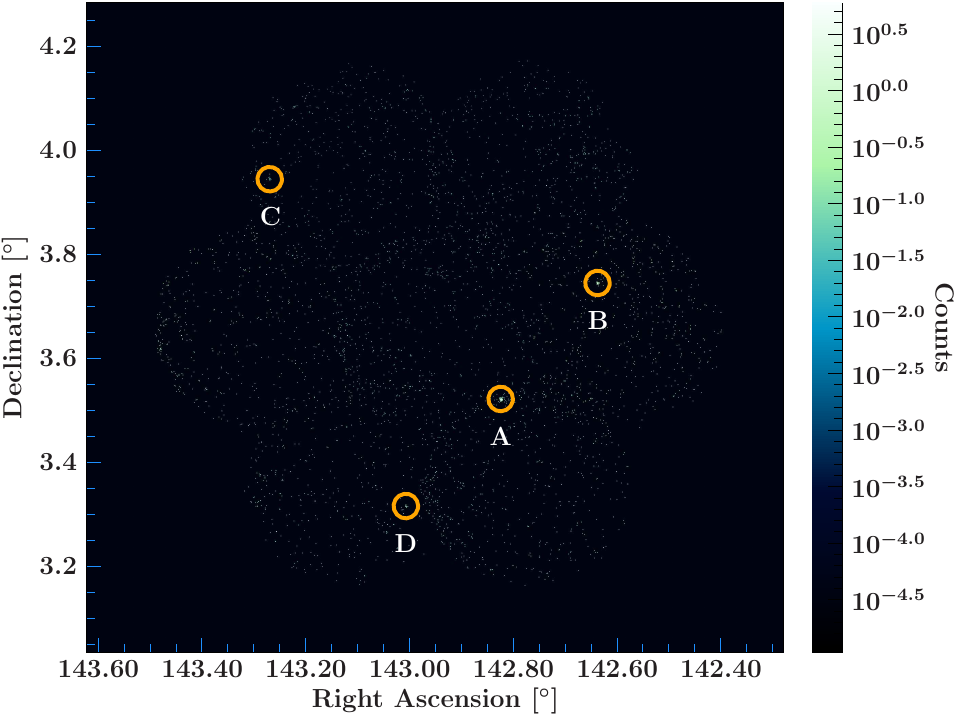}
    \caption{\Swift/XRT mosaic image with a 7-point tiling observation showing the detections of Sources A, B, C, and D.}
    \label{fig:xrt-mosaic}
\end{figure}

\begin{figure}
    \centering
    \includegraphics[width=0.5\textwidth]{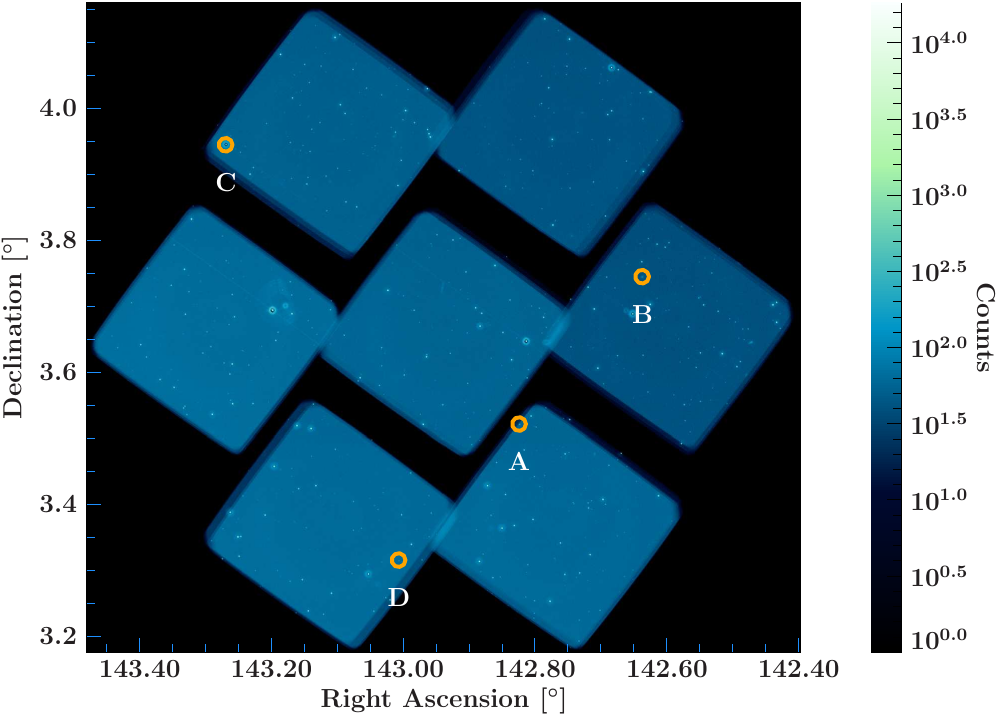}
    \caption{\Swift/UVOT mosaic image with a 7-point tiling observation showing the detections of Sources A, B, C, and D.}
    \label{fig:uvot-mosaic}
\end{figure}

The initial \Swift/XRT follow-up observations identified a total of 9 possible sources \citep{AMONAlert}. This included several spurious detections at low significance. All 9 sources are listed in Table~\ref{tab:srcs}. The numbers listed in the table are the ones we refer to in this Section.
The \Swift/XRT tiling observations are shown in Fig.~\ref{fig:xrt-mosaic} and the simultaneous \Swift/UVOT tiling observations are shown in Fig.~\ref{fig:uvot-mosaic}.

\begin{table*}[tb]
    \centering
    \caption{All \Swift/XRT detections listed by \citet{AMONAlert}. The number indicates the source number in the GCN, with their respective right ascension and declination [J2000.0], as well as the X-ray flux determined by XRT. The last three columns list any matches to known sources, a classification, as well as the letter given to the source in this work. Spurious detections have been greyed out, while source excluded due to their source type have been greyed out in a lighter grey.}
    \begin{tabular}{c|cccllc}
     Number & R.A. & Dec. & Flux & ID & Class.& this work\\
      & [$^\circ$] & [$^\circ$] & [erg\,s$^{-1}$\,cm$^{-2}$] & & & \\
     \hline
        1 & 142.82418 & 3.52179 & $(1.72\pm0.19)\times 10^{-12}$ & 1RXS\,J093117.6+033146 & Radio-Quiet AGN  & A  \\
        \cellcolor{gray!25}2 & \cellcolor{gray!25}142.75103 &\cellcolor{gray!25} 3.50071 & \cellcolor{gray!25}$10^{+7}_{-5} \times 10^{-14}$ &\cellcolor{gray!25}Spurious & \cellcolor{gray!25}&\cellcolor{gray!25} -\\
        3 & 142.63746 & 3.74534 & $8.5^{+1.8}_{-1.6} \times 10^{-13}$ & 2MASX\,J09303302+0344432 & Seyfert I & B\\
        \cellcolor{gray!15}4 & \cellcolor{gray!15}143.26942 & \cellcolor{gray!15}3.94512 & \cellcolor{gray!15}$3.8^{+1.3}_{-1.1}\times 10^{-13}$ & \cellcolor{gray!15}1RXS\,J093305.7+035648 & \cellcolor{gray!15}Star &\cellcolor{gray!15} C\\
        \cellcolor{gray!25}5 & \cellcolor{gray!25}142.65130 &\cellcolor{gray!25} 3.52369 & \cellcolor{gray!25}$3.34^{+1.22}_{-0.98}\times 10^{-13}$ & \cellcolor{gray!25}Spurious & \cellcolor{gray!25} &\cellcolor{gray!25} -\\
        6 & 143.00700 & 3.31602 & $3.34^{+1.22}_{-0.98}\times 10^{-13}$ & QSO\,J0932+0318 &  AGN &  D \\
        \cellcolor{gray!25}7 & \cellcolor{gray!25}142.74159 & \cellcolor{gray!25}3.53083 & \cellcolor{gray!25}$8^{+5}_{-4}\times 10^{-14}$ & \cellcolor{gray!25}Spurious &\cellcolor{gray!25}  & \cellcolor{gray!25}- \\
        \cellcolor{gray!15}8 & \cellcolor{gray!15}143.22296 & \cellcolor{gray!15}4.08343 & \cellcolor{gray!15}$2.31^{+1.10}_{-0.84}\times 10^{-13}$ &\cellcolor{gray!15}Unknown &\cellcolor{gray!15} & \cellcolor{gray!15}-\\
        \cellcolor{gray!15}&\cellcolor{gray!15}&\cellcolor{gray!15}&\cellcolor{gray!15}&\cellcolor{gray!15}likely spurious&\cellcolor{gray!15}&\cellcolor{gray!15}\\
        \cellcolor{gray!25}9 & \cellcolor{gray!25}142.74333 & \cellcolor{gray!25}3.80925 & \cellcolor{gray!25}$10^{+9}_{-6} \times 10^{-14}$ & \cellcolor{gray!25}Spurious & \cellcolor{gray!25} & \cellcolor{gray!25}-\\
    \end{tabular}
    \label{tab:srcs}
\end{table*}

While a total of 9 sources were detected and listed by \citet{AMONAlert}, four of these detections do not correspond to a known source, and were listed as likely spurious detections due low signal-to-noise ratios below 3$\sigma$. Source \#8 is at a low flux with 6 detected counts, near the edge of the detector. There is no known X-ray catalogue source in a 60$^{\prime\prime}$ radius. We consider it a spurious detection.
We exclude these sources from any further analysis due to their low statistical significance.
The four remaining sources are labelled A, B, C, and D, and correspond to previously detected sources.

Source C, which corresponds to 1RXS\,J093305.7+035648 has been identified as a star capable of producing X-rays, TYC 238-774-1 \citep{ROSATstars}.
However, stars produce neutrinos via fusion, and release MeV neutrinos \citep{nu_stars}. They cannot produce TeV neutrinos, and we exclude the star from our list of potential neutrino counterpart sources.

\begin{figure}
    \centering
    \includegraphics[width=0.5\textwidth]{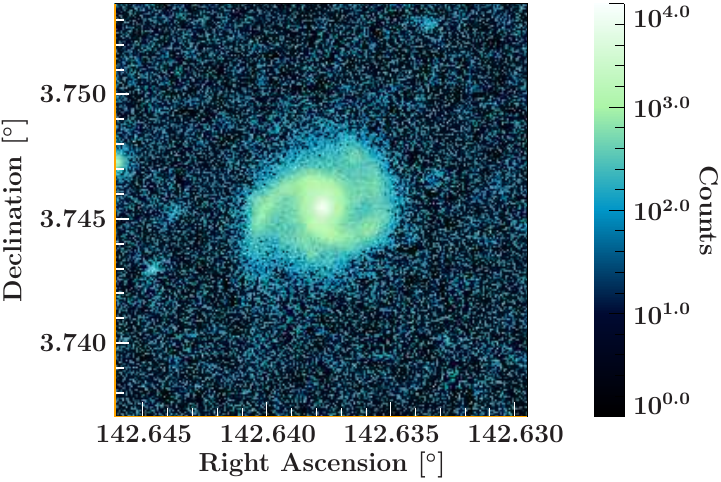}
    \caption{PanSTARRS image of Source B,  2MASX\,J09303302+0344432.}
    \label{fig:srcB-pan}
\end{figure}

Three potential sources remain: Source 1, ``A'', also known as 1RXS\,J093117.6+033146, which is a reasonably bright AGN; Source 3, ``B'', also known as 2MASX\,J09303302+0344432, which is Seyfert type I galaxy; and Source 6, ``D'', which is an AGN of unknown type.
It is worth noting that three likely radio-quiet AGN are the only catalogued X-ray bright sources in the uncertainty region of the neutrino event covered by \Swift/XRT tiling observations.
It is impossible to identify with absolute certainty which of these three AGN (if any) emitted the neutrino event, but we discuss a statistical approach in Sect.~\ref{sec:stat}.
Source B is a Seyfert I galaxy, 2MASX\,J09303302+0344432. A PanSTARRS view of the galaxy is shown in Fig.~\ref{fig:srcB-pan} \citep{panstarrs,panstarrs2,panstarrs3}. 
In this work, we argue that a higher flux in the X-ray band corresponds to a potentially higher neutrino flux, and we therefore focus our efforts on Source \#1 ``A'',\rxs. It is brighter than source \#3 (B), and \#6 (D) by a factor of $\sim$2 and $\sim$5, respectively. We obtained follow-up data of A with \nus and the VLA to identify its source type, and to calculate its estimated neutrino flux.

\subsubsection{Source A}
Source 1, ``A'', 1RXS\,J093117.6+033146, is a radio-quiet AGN at coordinates R.A.$=142\fdg 82419662563$, Dec$=3\fdg52209325452$ \citep{gaia} at a spectroscopic redshift of $z=0.20495$ \citep{sdss}.

\subsubsection{Radio Loudness}
\label{subsec:rl}
No definitive classification exists for \rxs, except for its general
QSO spectroscopic classification in the SDSS sample \citep{Albareti2015}.
Its detection in the radio band with the VLA allows us to determine
whether it is a radio-loud AGN. The standard radio loudness is defined as

\begin{equation}
 \label{eq:rl}
R = \frac{f_{5\,\mathrm{GHz}}}{f_{2500\text{\normalfont{\AA}}}} = \dfrac{0.089-0.9\,\mathrm{mJy}}{0.1-0.2\,\mathrm{mJy}}\simeq 0.1\qquad ,
\end{equation}
where $f$ refers to the differential fluxes in the radio band at 5\,GHz,
and the flux in the UV band at 2500\,\text{\normalfont{\AA}}\xspace \citep{Katgert1973,Condon1980,Kellerman1989,Stocke1992}.
Using an estimated range of fluxes from our VLA flux measurements and an estimate from the UVOT data as well as the SED model to predict the 2500\,\text{\normalfont{\AA}}\xspace flux, we obtain a very low
radio-loudness of $R\sim0.1$.
This confirms that \rxs is likely a Seyfert type galaxy, though it is unclear whether it is jetted.

\subsection{Other potential counterparts}
\label{subsec:other-srcs}
In order to be able to understand the statistical significance of our possible association of a source with a neutrino event, we study X-ray sources outside the detection area of \Swift/XRT, but within the neutrino uncertainty area.
In the 50\% uncertainty region of the neutrino event, two sources are listed in the 2RXS catalogue \citep{2RXS}. 
Sources ``B'' (2MASX J09303302+0344432) and ``D'' (QSO J0932+0318), which were detected in the \Swift/XRT tiling were not detected by ROSAT. No additional catalogue sources are known. Within the 90\% uncertainty region a total of nine sources are known (see Table~\ref{tab:rosat90}). Of these 9 detections, two are the ``A'' and ``C'' sources detected by XRT. Only one catalogue source is found within the XRT area but was not detected by XRT, likely due to a low flux state. All ROSAT-only sources are labelled with small Roman numerals. Source iii detected in ROSAT is another X-ray detected star, which is excluded from the analysis. This leaves an additional 5 X-ray sources inside the 90\% uncertainty area, but not inside the XRT covered area.
We analyse the ROSAT detected sources and calculate the expected neutrino numbers of these sources in Sect.~\ref{subsec:nunum-other}. This information is used in Sect.~\ref{sec:stat} to determine which source is the most likely counterpart.
The brightest listed source 1RXS J093107.0+030447 is the star HD\,82667, while the second brightest source is 1RXS\,J093433.1+033316, which likely corresponds to the star HD\,82821; as well as the X-ray emitting star 1RXS J093305.7+035648 (TYC 238-774-1). 
The brightest likely neutrino source identified by ROSAT is source A, \rxs. Fainter listed sources include 1RXS J093505.3+034510 (unknown type), 1RXS J093433.3+030631 (radio galaxy PMN J0934+0305), 1RXS J092918.3+041917 (Seyfert I galaxy 2MASS J09291833+0419332), 1RXS J093003.0+035702 (unknown type), and 2RXS J093149.8+034632 (SDSS J093150.51+034629.6, QSO). 
We note that these include only (with the exception of the stars), AGN, including mostly radio-quiet AGN, with the exception of the radio galaxy.

\begin{table*}\scriptsize
    \centering
    \caption{ROSAT detections in the 90\% uncertainty region}
    \begin{tabular}{cccc|cc|ll|ll}
    ID  & s & Type & 50\% & 1RXS name & 2RXS name & $\alpha_{\mathrm{J2000}}$ &  $\delta_{\mathrm{J2000}}$ & $F_{\mathrm{2RXS},0.1-2.4\,\mathrm{keV}
    }$ & $F_{\mathrm{Swift},2-10\,\mathrm{keV}}$\\
     & $[^\prime]$ & & & & &  [$^\circ$] & [$^\circ$] & $10^{-13}[$erg/s/cm$^2]$ & $10^{-12}[$erg/s/cm$^2]$ \\
    \hline
   A & 10.9 & QSO & y &  J093117.6+033146 &  J093117.5+033146 & 142.823250 & 3.529611 &  $5.4\pm1.6$ & $1.72\pm0.19$ \\
   i & 7.0 & QSO &  y  & & J093149.8+034632 &   142.957708 & 3.775583 & $2.7\pm1.2$ & -\\
   \hline
     \cellcolor{gray!25}C &  \cellcolor{gray!25}25.9 &  \cellcolor{gray!25}Star &  \cellcolor{gray!25}n &  \cellcolor{gray!25} J093305.7+035648 &  \cellcolor{gray!25}J093305.6+035648 &  \cellcolor{gray!25}143.273583 & \cellcolor{gray!25} 3.946861 &  \cellcolor{gray!25}$3.3\pm1.4$ &  \cellcolor{gray!25}\\
     ii & 31.5 & ? & n & J093003.0+035702 & J093003.0+035702 & 142.512833 & 3.950611 & $3.3\pm1.3$ & \\
      \cellcolor{gray!25} iii &  \cellcolor{gray!25}36.29 &  \cellcolor{gray!25}Star  & \cellcolor{gray!25} n & \cellcolor{gray!25}J093107.0+030447&  \cellcolor{gray!25}J093106.8+030447 & \cellcolor{gray!25} 142.775802 & \cellcolor{gray!25} 3.079790 &  \cellcolor{gray!25} $18.8\pm2.8$ &  \cellcolor{gray!25}\\
     iv &   41.65 & ? & n &  J093433.1+033316 & J093433.0+033317 & 143.6379 & 3.5544 & $6.8\pm1.9$ & \\
      v & 49.33 & ? & n & J093505.3+034510 & J093504.5+034458 & 143.772085 & 3.752775  & $4.8\pm1.8$ & \\
      
      vi & 52.85 & Radio G & n& J093433.3+030631 & J093433.0+030628 & 143.627857 & 3.095922 & $4.7\pm1.7$ & \\

     vii & 54.47 & Sey 1 & n & J092918.3+041917 &  J092918.5+041919 & 142.326459 & 4.325938 & $4.4\pm1.5$ & \\
    \end{tabular}
    \label{tab:rosat90}
\end{table*}

\subsection{Broadband SED}
\label{subsec:sed_results}
\begin{figure*}
    \centering
    \includegraphics[width=\textwidth]{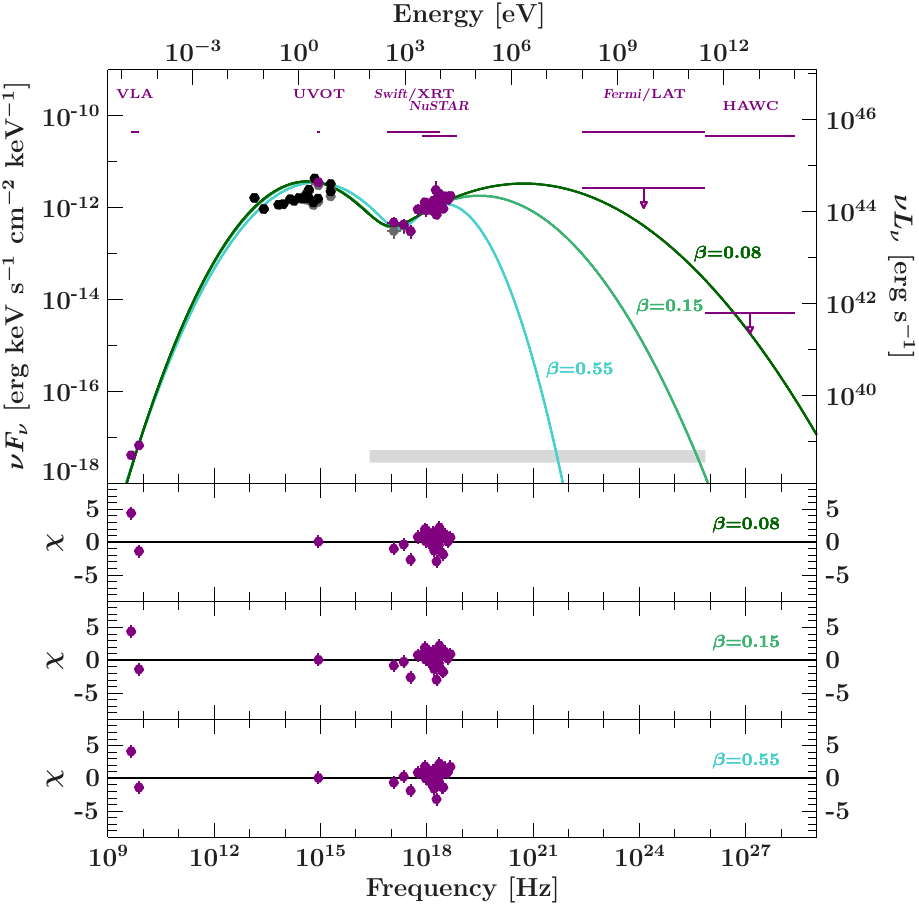}
    \caption{Multiwavelength SED of \rxs. Quasi-simultaneous data is
      shown in purple, while archival data is shown in black. Absorbed
    and reddened (raw) data are shown in gray. The energy band used for the
    integration of the energy flux is shown in gray at the bottom
    right side. Multiple possible parabola models are shown in three
    colours: dark green, green and cyan. Due to the lack of a 
    $\gamma$-ray detection, it is  unclear which model best describes the source. 
   The bottom panel shows the
    residuals for the purple model ($\beta=0.15$). All three models
    fit the X-ray data equally well, with negligible differences.}
    \label{fig:sed}
\end{figure*}
\begin{figure*}[bth]
    \centering
    \includegraphics[width=0.45\textwidth]{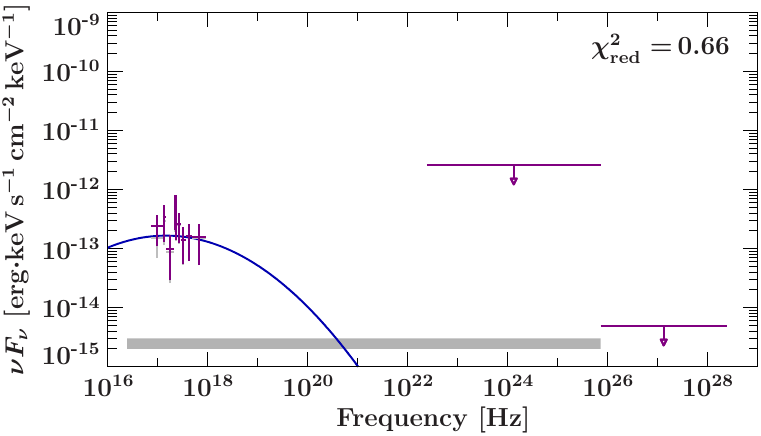}\qquad%
       \includegraphics[width=0.45\textwidth]{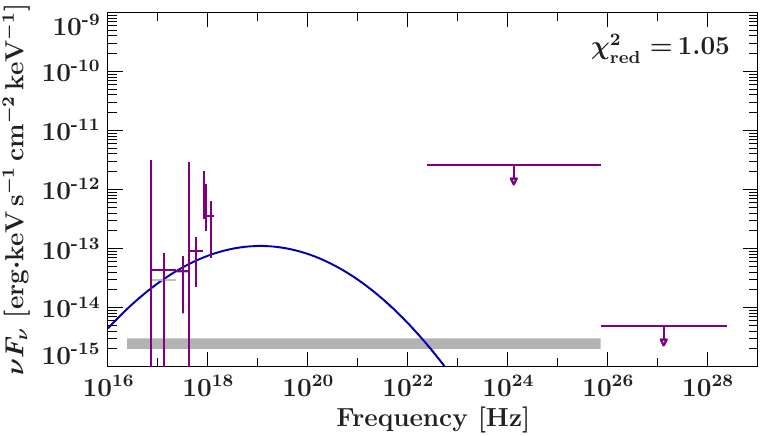}
    \caption{SEDs for Source B, 2MASX J09303302+0344432 (left) and Source D, QSO J0932+0318 (right) with the \Swift/XRT data and the \Fermi/LAT and HAWC upper limits. A best fit model is shown in blue.}
    \label{fig:other-seds-xrt}
\end{figure*}
\begin{figure*}[bth]
    \centering
    \includegraphics[width=0.45\textwidth]{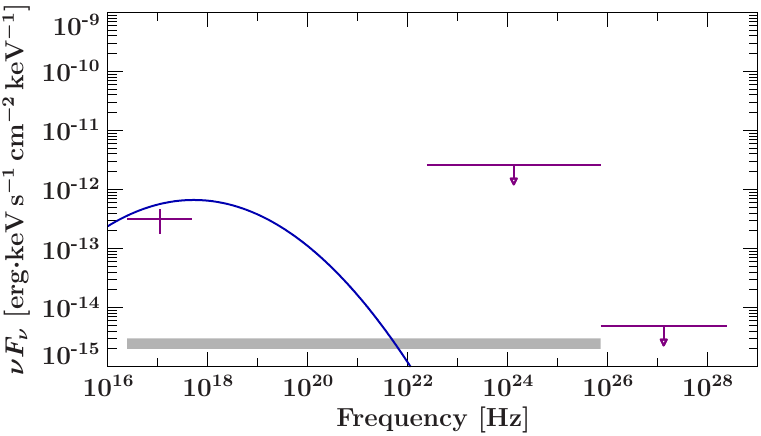}\qquad%
    \includegraphics[width=0.45\textwidth]{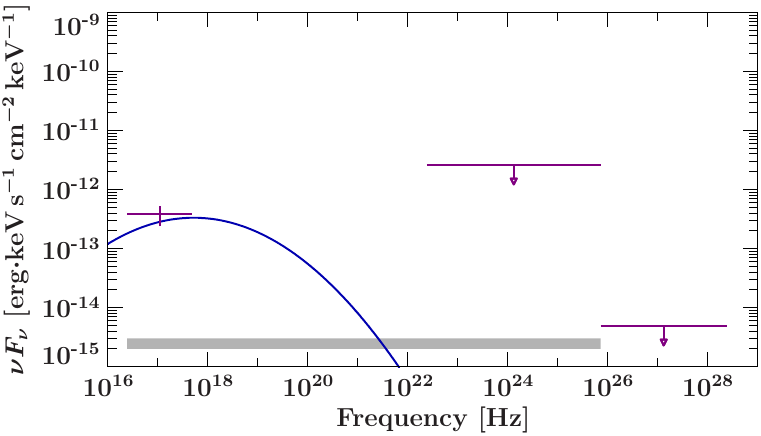}\\
    \includegraphics[width=0.45\textwidth]{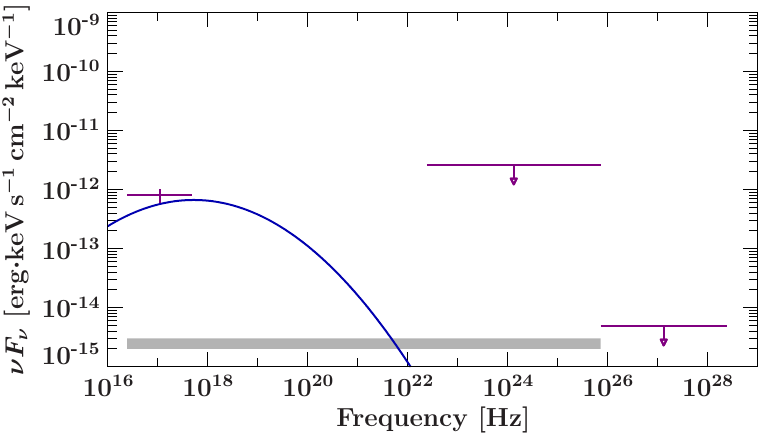}\qquad%
    \includegraphics[width=0.45\textwidth]{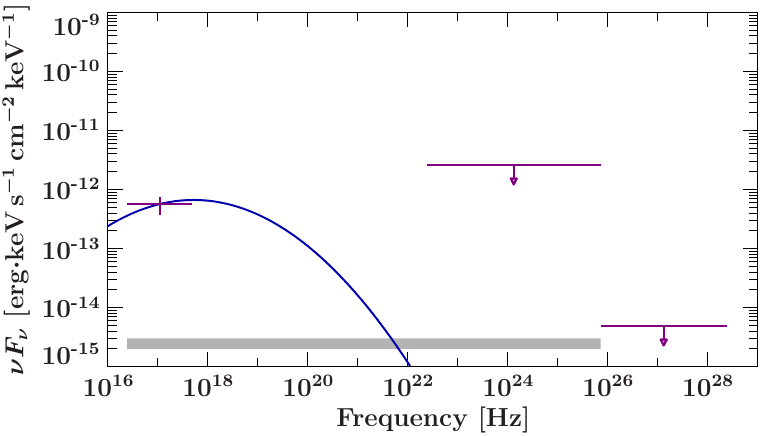}\\
        \includegraphics[width=0.45\textwidth]{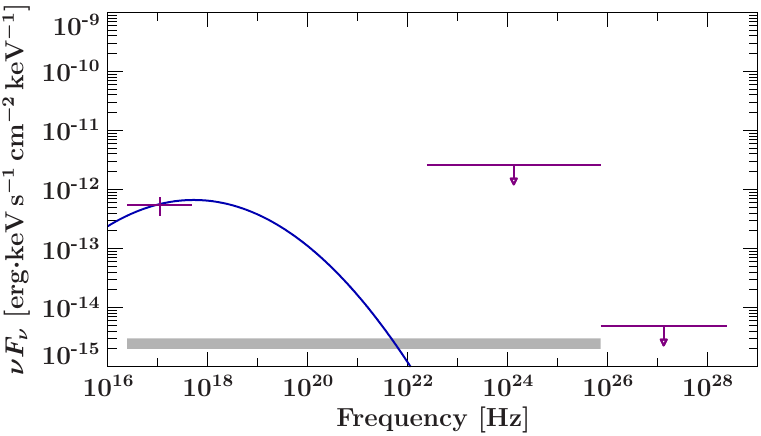}\qquad%
    \includegraphics[width=0.45\textwidth]{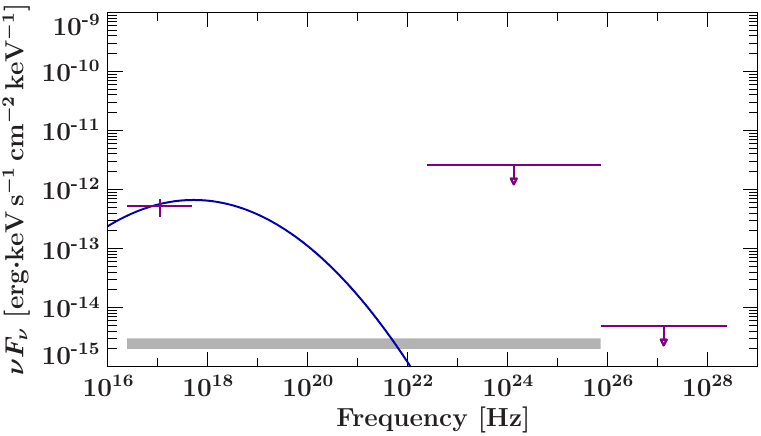}\\
    \caption{SEDs for Source i (top left) and Source ii (right) with the ROSAT data and the \Fermi/LAT and HAWC upper limits. A best fit model is shown in blue.
    The second row shows the SEDs for Sources iv (left) and v (right). The bottom row shows the SEDs for sources vi (left) and vii (right).
    }
    \label{fig:other-seds-rosat}
\end{figure*}

We generate quasi-simultaneous broadband SEDs of \rxs (see
Fig.~\ref{fig:sed}) and the other potential counterparts (see Fig.~\ref{fig:other-seds-xrt} for the SEDs of sources B and D and Fig.~\ref{fig:other-seds-rosat} for the SEDs to the ROSAT-only detected sources). The observation dates range from same day
observations(\Fermi/LAT and \Swift/XRT, both observed on June 15th, 2020) to
being taken 2 months after the neutrino detection (VLA data
taken on August 25th, 2020).  We include flux upper limits in the $\gamma$-ray
band from \Fermi/LAT and HAWC, which are constraining for the model of the high-energy peak.
In order to estimate a possible range of neutrino numbers we fit three
different models with a fixed curvature of $\beta=0.08$ (blue),
$\beta=0.15$ (purple) and $\beta=0.55$ (green). The X-ray data alone
is unable to distinguish between these models, which can be seen in the residual panels.
We use all three models to estimate a range of neutrino numbers.
Information about the three models and the model statistics is listed in Table~\ref{tab:fit-sed}.
\begin{table*}
     \centering
     \caption{This presents  best fit results. The three columns represent the three different models applied to the SED in Fig.~\ref{fig:sed}. The first section shows the best fit parameters for the three different models. The first four refer to the fit parameters of the synchrotron parabola including the normalization $N_1$, the slope $\alpha_1$ and the curvature $\beta_1$ at the pivot energy $E_{p,1}$. For the second parabola, the normalization $N_2$ is given in units of $10^{-7}$. The slope $\alpha_2$ at the pivot energy $E_2$ is also listed. As the curvature $\beta_2$ defined the fit, it is listed at the top of the table. Below the best fit parameters, we list the $\chi^2$ and reduced $\chi^2$ values. $\Phi_\gamma$ lists the integrated photon flux of the high-energy peak which has been used to calculate neutrino numbers, which are given in Table~\ref{tab:nunum}.}
    \begin{tabular}{c | ccc}
    \hline
      $\beta_{2}$ & 0.08 & 0.15 & 0.55 \\
      \hline
      $N_{1}$ & $0.28^{+0.13}_{-0.14}$ & $0.30^{+0.13}_{-0.14}$ & $0.40^{+0.10}_{-0.13}$\\
      $\alpha_{1}$ & $2.72^{+0.31}_{-0.17}$  & $2.69^{+0.29}_{-0.16}$ & $2.56^{+0.17}_{-0.10}$\\
      $\beta_{1}$ & $0.251^{+0.042}_{-0.023}$ & $0.247^{+0.039}_{-0.021}$ & $0.229^{+0.024}_{-0.014}$\\
      $E_{p,1}$ & 0.05 & 0.05 & 0.05 \\
      $N_{2}$ [$10^{-7}$] & $1.6^{+0.6}_{-0.4}\times10^{-7}$ & $1.3^{+0.5}_{-0.4}$ & $0.31^{+0.15}_{-0.10}$\\
      $\alpha_{2}$ &$1.75^{+0.10}_{-0.11}$ &$1.93^{+0.11}_{-0.12}$ & $2.92^{+0.14}_{-0.15}$\\
      $E_{p,2}$ & 100 &100 & 100\\
      \hline
    $\chi^2$ & 73.63 & 71.96 & 69.52 \\
    $\chi^2_{\mathrm{red}}$ & 2.17 & 2.12 & 2.05\\
      \hline
    $\Phi_\gamma$ [$10^{-11}$] & $3.96\pm0.24$ & $1.42\pm0.09$ & $0.45\pm0.03$\\
    \hline
    \end{tabular}
    \label{tab:fit-sed}
   \end{table*}

\subsection{Neutrino Numbers}
\subsubsection{Neutrino Numbers of Source A}
\label{subsec:nunumres}
We calculate the maximum and scaled neutrino numbers as described in
Section.~\ref{subsec:nunum}. 
   \begin{table*}
     \centering
     \caption{This table lists the maximum neutrino number, a 30\% scaled neutrino number (and their respective Poisson probabilities $P$ and an unrealistic delta-function model in the last column. These numbers are provided for two different neutrino slopes, $\Gamma_\nu=1.9$ and $\Gamma_\nu=2.7$}
    \begin{tabular}{c| c | ccc}
    \hline
     & $\beta_{2}$ & 0.08 & 0.15 & 0.55 \\
      \hline
    \multirow{5}{*}{\rotatebox[origin=c]{90}{$\Gamma_\nu=1.9$}}& $N_{\nu_{\mu,\mathrm{max}}}$ & 0.408 & 0.162 & 0.057 \\
     & $P$ & 0.27 & 0.14 & 0.10 \\
    \cline{2-5}
    & $N_{\nu_{\mu},\mathrm{scaled}}$ & 0.12 & 0.05 & 0.017\\
    & $P$ & 0.11 & 0.05 & 0.02 \\
    \cline{2-5}
    & $N_{\nu,\delta}$ & 0.55 & 0.22 & 0.08 \\
\hline
\hline
    \multirow{5}{*}{\rotatebox[origin=c]{90}{$\Gamma_\nu=2.89$}}& $N_{\nu_{\mu,\mathrm{max}}}$ & 0.471 & 0.187 & 0.066  \\
     & $P$ & 0.29 & 0.16 & 0.06 \\
    \cline{2-5}
    & $N_{\nu_{\mu},\mathrm{scaled}}$ & 0.14 & 0.06 & 0.02 \\
    & $P$ & 0.12 & 0.05 & 0.02 \\
    \cline{2-5}
    & $N_{\nu,\delta}$ & 0.55 & 0.22 & 0.08 \\
\hline
    \end{tabular}
    \label{tab:nunum}
   \end{table*}

The neutrino spectrum needed to calculate the neutrino numbers is uncertain due to a lack of sufficient neutrinos for one individual extragalactic source. We apply the simplest model, a powerlaw. While the ``photon'' index of the neutrino spectrum of an individual source is also unknown, we apply a hard index of $\Gamma_\nu=1.9$, as well as the soft index that is a best fit to the IceCube neutrino data $\Gamma_nu=2.89$ \citep{IceCube75}.
Results for all three models with both indices are listed in Table~\ref{tab:nunum}.

\subsubsection{Neutrino Numbers of other sources}
\label{subsec:nunum-other}
We apply the same approach to the other sources to identify possible neutrino numbers (``weights'') for the other XRT-detected and the ROSAT-detected sources. Results are listed in Table~\ref{tab:swift_sources}. The additional SEDs for this approach can be found in Fig.~\ref{fig:other-seds-rosat}.

\section{Discussion}
\label{sec:disc}
\subsection{Neutrino uncertainty area}
\label{subsec:nuunc}
Muon track events are well localized with uncertainty radii $\lesssim
1^\circ$. This can still correspond to large uncertainty area that
cannot easily be covered with $\le 10$ tiling observations in the
X-ray band. The 50\% uncertainty radius of IceCube-200615A is
21.5$^{\prime}$, while the 90\% uncertainty radius is $55.2^{\prime}$.
This corresponds to a difference in area of more than a factor of 5.
With a 7-point tiling \Swift/XRT observation we cover more than double
the 50\% uncertainty region but only 37.6\% of the 90\% uncertainty
region. Probability drops with increasing distance from the best fit
coordinates.
However, it is possible that we have not observed the actual
neutrino counterpart, due to it being located outside of the observed \Swift/XRT tiled area.
The second ROSAT catalogue lists an additional 5 sources in the 90\% uncertainty
region (see Sect.~\ref{subsec:other-srcs}).
We note that AGN are the only potential X-ray-bright counterparts identified by our observations of IceCube-200615A, based on the detections of the fainter sources B and D and an analysis of the ROSAT catalogue data.

\subsection{Neutrino Numbers}
Table~\ref{tab:nunum} lists the resulting neutrino numbers calculated following Section~\ref{subsec:nunum} and Section~\ref{subsec:scale}. The resulting numbers for Source A are in the range of 0.017 (scaled) -- 0.471 (maximum neutrino number). These numbers should be considered upper limits as we assume that all of the high-energy peak (or 30\% for the scaled numbers) are produced in hadronic processes which is unlikely.
While the neutrino numbers are below 1, we do not expect the counterparts to necessarily reach $\sim 1$ expected neutrino count \citep{Krauss2018}. If all reasonably bright radio-quiet AGN had an expected neutrino number $\gtrsim 1$, we'd expect to detect a much higher flux of astrophysical neutrinos from IceCube.\\
We therefore also provide the Poisson probability of detecting exactly one neutrino event under the assumption that the average neutrino rate is given by the maximum neutrino number. This yields probabilities of $\sim$ 2\% -- 27\%. This is consistent with our observations of only detecting promising neutrino events rarely.

\section{Statistical analysis}
\label{sec:stat}
In this section, we present a statistical analysis for the sources identified in Sect.~\ref{subsec:srcs} via \Swift/XRT tiling observations. We also consider ROSAT-detected X-ray sources that are discussed in Sect.~\ref{subsec:other-srcs}. 
\subsection{Introduction}
We present a probabilistic model to evaluate the relative strength of candidate source locations for a specific neutrino detection event.  As part of the initial alert, we receive a position in the sky, a 50\% radius of detection, and a 90\% radius of detection. Within this area, \Swift/XRT and ROSAT identify $K=9$ possible point locations consistent with a possible neutrino emission. The goal of this model is to quantitatively assess the candidacy of possible sources, not to identify a definitive source.

While each point location identified within the \Swift/XRT window is the possible source of the neutrino event, we must also consider the possibility that either that of the 63,141 relevant candidate objects in the ROSAT catalogue may be the source, or that the source may originate from a previously undetected object. 
Our analysis constructs a probabilistic model to calibrate the relative support of these candidate sources in order to calculate a Bayes factor for each possible source.
This provides an exhaustive ranking of all possible candidate sources and concludes that Source A has overwhelming support relative to all other possibilities. 

We assume that the  putative source location identified by IceCube is modelled by a bivariate Gaussian distribution centred at the detection centre. 
The detection process places a radius of 21.5$^\prime$ for a 50\% probability and 55.2$^\prime$ for a 90\% probability.
The probability of each point location being the source is then given by its radius from the detection centre. 
As the detection distribution is a bivariate Gaussian, the distribution of the radius from the centre follows a Rayleigh distribution with variance parameter $\sigma$.
Due to slight inconsistencies in the specification of the detection distribution, we find that the variance parameter $\sigma \in [18.26,25.72]$.

We assume that the strength of the observed flux for each candidate object is proportional to its probability of being the neutrino source. 
The probabilistic observation model then combines two possible pieces of data for each possible source location: the angular distance of each object relative to the identified source origin, and the observed flux of each object.  
Below, we describe the two probability models for each type of data, how they are integrated into a unified Bayesian framework and inference is performed, and finally a summary of the relative support the model provides for each candidate source.

\subsection{Data}
Our analysis includes four distinct data sets: the initial IceCube detection; the follow-up \Swift/XRT observation process; the ROSAT catalogue of candidate sources \citep{2RXS}; and the eROSITA catalogue of candidate sources \citep[eRASS; ][]{erass}. 
We label these $D_i$, $D_S$, $D_R$ and $D_E$, respectively. As the eRASS catalogue is limited to the Western half of the sky, it does not include the region of the sky that IceCube identifies as the putative location of the neutrino event, its contribution to the combined probability model is indirect, providing information on the number of unobserved candidate objects. We consider the particular contributions of each data source below.

\subsubsection{IceCube data}
The IceCube detector provides two pieces of information upon detection of a neutrino event that are integrated into the overall model: the probability that the observed event is celestial in origin and the putative origin of the neutrino together with a variance.
As our analysis is conditional upon the assumption that the neutrino is celestial in origin, the first piece of information does not contribute to our model. 
The second piece of information -- the detection centre -- is integrated into the analysis as a recentring of all of the candidate sources in degrees from this origin point (angular distance), and so effectively `disappears' into the data.
The only information remaining included in the model is the associated uncertainty for the variance parameter $\sigma$.

\subsubsection{\Swift/XRT data}
The \Swift/XRT observation data provide us with the observed flux and the angular distance from the detection centre for four possible sources. 
For notation, we write $D_S=\{ d_1=(r_1,f_1),\cdots, d_K=(r_K,f_K) \}$, where $r_i$ is the angular distance from the detection centre for point location $i$ and $f_i$ is the scaled neutrino flux for that point location. 
We assume that data values are independent conditional on point location and that $r_i$ and $f_i$ are conditionally independent given whether or not the point location is the source. The values for these possible sources are listed in Table~\ref{tab:swift_sources}.

\begin{table}[!ht]
\centering
\caption{Summary of point sources identified with \Swift/XRT (letters) and ROSAT (Roman numbers). The last column gives whether the point location was detected by \Swift/XRT (XRT), both the survey and secondary observation (both), not detected by ROSAT despite being in the \Swift/XRT field of view (undetected), and only in the survey (ROSAT).
This Table lists the maximum neutrino number, a 30\% scaled neutrino number for all sources. These numbers are used in Sect.~\ref{sec:stat}. These numbers assume a neutrino slope of $\Gamma=2.89$ and $\beta=0.15$.
}
\begin{tabular}{cccc}
Source & Angular distance & weight & detection \\ 
\hline A & 0.1865 & 0.06 & XRT \\
B & 0.3231 & 0.00384 & XRT \\ 
D & 0.3683 & 0.00341 & both \\
i & 0.115839 & 0.00222 & undetected \\
ii & 0.524145 & 0.00266 & ROSAT \\
iv & 0.694611 & 0.00556 & ROSAT \\ 
v & 0.825595 & 0.00389 & ROSAT \\
vi & 0.880951 & 0.00382 & ROSAT \\
vii & 0.911256 & 0.00356 & ROSAT \\
\end{tabular}
\label{tab:swift_sources}
\end{table}

\subsubsection{ROSAT data}
To consider all possible candidate sources, we include all 63,141 possible sources from the ROSAT survey catalogue, using a flux cut-off of $1.8\times10^{-14}$\,erg/s/cm$^2$.
For notation, we use a similar framework to that for the \Swift/XRT data: $D_R=\{ d_1=(r_1,\lambda_1),\cdots, d_K=(r_K,\lambda_K) \}$ where $K=63,141$  and $r_i$ have the same meaning as before.
We write $\lambda_i$ for the observed flux to make clear that the ROSAT observation is calibrated differently from the \Swift/XRT data and so our model must account for this discrepancy.

\subsubsection{eROSITA catalogue (eRASS)}
While the eRASS catalogue does not include the neutrino uncertainty region identified by the IceCube detector, it provides important information on the total number of unobserved candidate sources in the sky.
We filter these data in the following fashion: first, we restrict the ROSAT catalogue to the region covered by the eRASS catalogue; second, we scan through the ROSAT and eRASS catalogues to identify the number of objects found in both catalogues (through angular distance matching) to get an estimate; 
finally, we record the total number of objects found in each catalogue, which we denote as $|R|$ and $|E|$, respectively, and the number of objects shared jointly in both catalogues, $|K|$.  We then write $D_E = (|R| = 63,141,|E| = 929,055, |K| = 35,397)$.

\subsection{Number of objects considered}
The total number of objects considered across all data sets is then the number of objects in the ROSAT catalogue $(|R|)$, the number identified with the \Swift/XRT detector and ROSAT ($|S|=9$), and the number of unobserved objects $M-|R|-|S|$, where $M$ is the total number of candidate objects. 
As we have only indirect information about $M$, we treat this as a random variable to be inferred by our model. 
Furthermore, as the unobserved objects necessarily have no observed flux or position, we can treat all unobserved objects as equivalent representatives of a single class. 
Thus, the total number of separate objects considered in the model is then $|S|+|R|+1 = 63,141+9+1= 63,151$.

\subsection{Model}
As mentioned above, the observational model consists of two separate components:
the location model that gives the likelihood of a particular point location being the source of the neutrino relative to the detection location; 
and the flux model that calibrates the likelihood of a source being the neutrino origin as a function of its strength.  
Both models provide a probability for each possible source, including unobserved objects, which we treat as a single class of objects in the model.
We combine the two models into a joint observational model using a latent variable formulation with a vector of augmented variables $\boldsymbol{z} = \{z_1,\cdots, z_{|S|+|R|+1} : z_i \in \{ 0,1\} \mbox{ and } \sum_{k=1}^{|S|+|R|+1} z_k = 1 \} $ that specify which of the candidate locations is the neutrino source, i.e., for which $k$ does $z_k=1$.
The inferential goal of the model is to find the marginal posterior distribution for $\boldsymbol{z}$ having integrated out the other forms of uncertainty in the model.

\subsubsection{Flux model} 
The flux model calibrates the relative contribution of the observed flux of a candidate source as evidence of being the putative neutrino source.
If the flux of a particular source is $f_i$ and is held to the be source of the neutrino (i.e. $z_i=1$) then the observational probability is given by a Poisson distribution not being zero, i.e., $ \mbox{POISSON}(>0|f_i) = 1-e^{-f_i}$.
If it is not the source, then probability is given by the probability of a Poisson distribution having zero count, $\mbox{POISSON}(0|f_i)=e^{-f_i}$.
We assume that the flux distribution of all candidate sources is given by a log-normal distribution with parameters 
$\mu$ and $\eta^2$.
The Poisson distribution is an approximation that only holds for brief periods of time. 
Over a longer time, the underlying Poisson process is  inhomogeneous, and better approximated as a negative binomial distribution. 
This allows for additional dispersion relative to that expected from the Poisson distribution, with the challenge that the negative binomial often has unstable parameter estimates. With the limited available ROSAT and \Swift data, we employ a Poisson distribution for computational stability.

As the ROSAT candidate sources are calibrated differently than the \Swift/XRT sources, we homogenize them in the following way: we find the percentile for each of the ROSAT sources in the data set, and then use the log-normal distribution with $\mu$ and $\eta$ to find comparable $f_i$ values for use in the flux model.
For unobserved sources, we take the expected value of the log-normal distribution to be the flux, which we denote as $\overline{f}$.  

\subsubsection{Location model}
The location model assumes that if a particular candidate is the neutrino source ($z_i=1$) then it follows a Rayleigh distribution with variance parameter $\sigma$.
If a particular candidate is not the source, then, assuming that the candidate sources are homogeneously distributed throughout the sky, it follows that the probability of observation is a function of the radius from the detection centre, $\frac{r_i}{180}$. Alternatively, a normal distribution could be constructed for this portion of the model, but numerical experiments show little improvement in fit at the cost of a more involved MCMC scheme so we employ the simpler approach. For the unobserved values, we take the position to be the average over all of the positions in the ROSAT catalogue. We denote this value as $\overline{r}$. Numerical experiments showed almost no change in the companion source identification between taking the average value and integrating over the uncertainty in the unobserved angular distances $r_i$.

\subsubsection{Unobserved sources model}
As the number of unobserved sources is unknown, we use a hypergeometric distribution to estimate the size of the entire population.
If the size of the total population is $M$ 
and we have made two surveys of size $|R|$ and $|E|$ that have overlap $|K|$, the probability of taking a value $M$ is given by:
$$\mathbb{P}(|K| | M, |R|,|E|) = \dfrac{ \binom{M-|E|}{|K|}  \cdot  {\binom{|E|}{|R|-K}}}{{\binom{M}{|R|}}} ~~\mbox{.}$$

\noindent The number of unobserved sources is then $M- |S|-|R|$. 

\subsubsection{Full observational model}
The full likelihood $ \Pi(D_i,D_S,D_R,D_E| \boldsymbol{z},  \mu, \eta, \sigma,M)$ is constructed by combining the flux model and the location model using the latent variable formulation.
As $D_i$ effectively ``disappears'' into the variance parameter $\sigma$ and $D_E$ contributes only to $M$, the observational model can then be reduced to $ \pi(D_S,D_R| \boldsymbol{z}, \mu,\eta, \sigma,M)$. \\

\noindent \textbf{If $z_i=1$ and $i \in S \cup R$}
\begin{equation}
\begin{split}
 \pi(D_S,D_R| \boldsymbol{z},  \mu, \eta, \sigma,M) \\
 =   \bigg\{  \mbox{POISSON}(>0|f_i)  \cdot \prod_{j \neq i}^{|S|+|R|} 
 \mbox{POISSON}(0|f_j)  \bigg\}  \\ \cdot \mbox{POISSON}(0|\overline{f})^M  \\
  \cdot \bigg\{  \prod_{i =1}^{|S|+|R|} \mbox{log-}\mathcal{N}(f_i | \mu,\eta^2)  \bigg\} \cdot  \mbox{log-}\mathcal{N}(\overline{f}_i | \mu,\eta^2)^M   \\
  \cdot \mbox{RAYLEIGH}(r_i | \sigma) \cdot \bigg\{  \prod_{j\neq i }^{|S|+|R|}  \cdot \frac{r_j}{180} \bigg\} \cdot \bigg( \frac{\overline{r}_i}{180} \bigg)^M\mbox{.}\\
\label{likelihood}
\end{split}
\end{equation}

\noindent \textbf{If $z_i=1$ and $i \notin S \cup R$}
\begin{equation}
\begin{split}
 \pi(D_S,D_R| \boldsymbol{z},  \mu, \eta, \sigma,M)  =  \bigg\{  \mbox{POISSON}(>0|\overline{f})  
 \\ \cdot \prod_{i=1}^{|S|+|R|} 
 \mbox{POISSON}(0|f_j)  \bigg\}  \cdot \mbox{POISSON}(0|\overline{f})^{M-1}  \\
  \cdot \bigg\{  \prod_{i =1}^{|S|+|R|} \mbox{log-}\mathcal{N}(f_i | \mu,\eta^2)  \bigg\} \cdot  \mbox{log-}\mathcal{N}(\overline{f}_i | \mu,\eta^2)^M \\
  \cdot \mbox{RAYLEIGH}(\overline{r}_i | \sigma) \cdot \bigg\{  \prod_{j \neq i }^{|S|+|R|}  \cdot \frac{r_j}{180} \bigg\} \cdot \bigg( \frac{\overline{r}_i}{180} \bigg)^{M-1}\mbox{.}
\end{split}
\end{equation}

\subsubsection{Prior specification}
\begin{figure}
\begin{center}
\includegraphics[width=\linewidth]{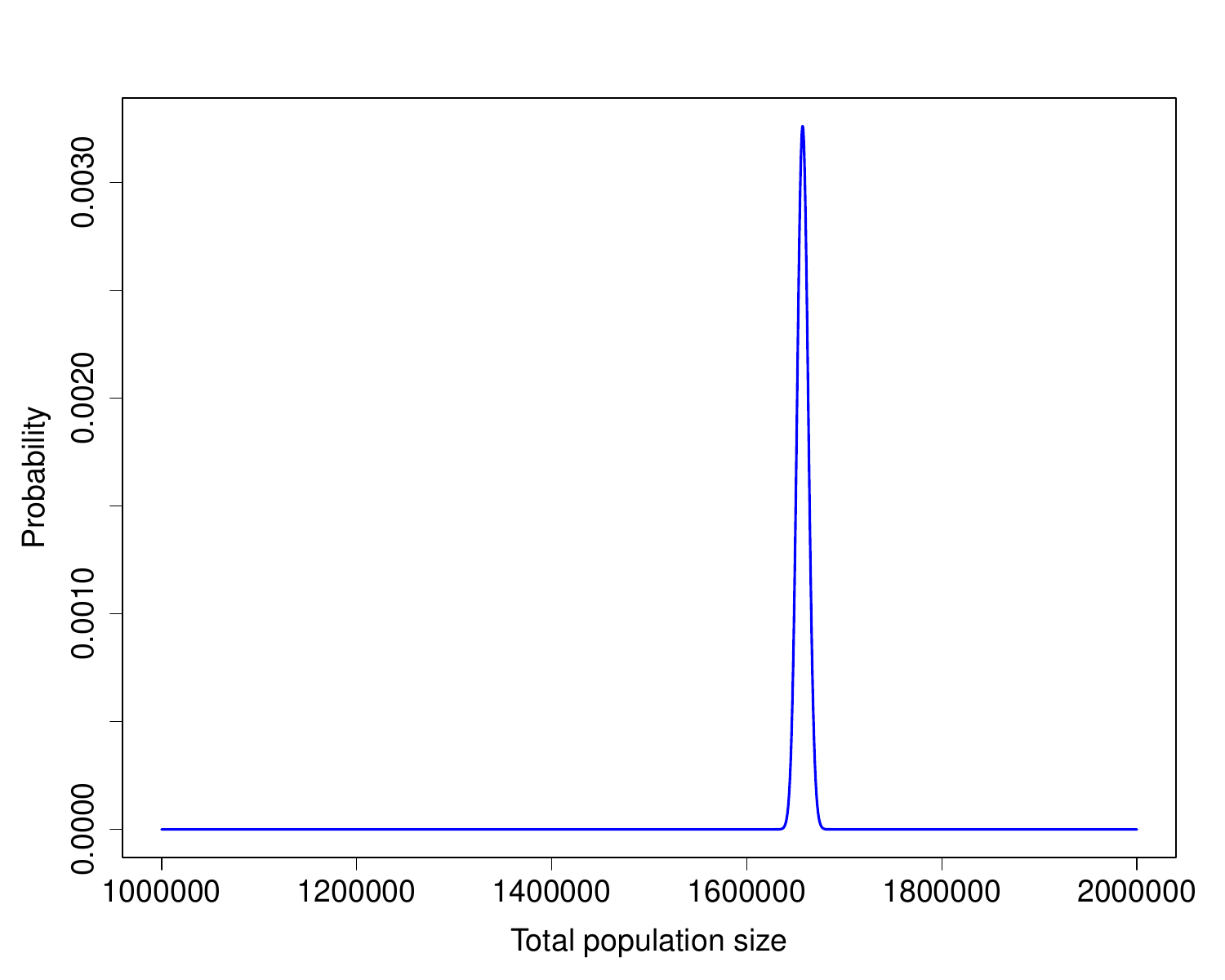}
\end{center}
\caption{Probability of different values of $M$ using a hypergeometric distribution. The peak lies at $M=1,657,244$.}
\label{hypergeo}
\end{figure}

To complete our model, we require prior distributions on the parameters.  
\begin{itemize}
\item $\boldsymbol{z}$ : \emph{A priori} any of the point locations may be the source with equal probability so $z_i=1$ occurs with probability $\frac{1}{M}$ and, correspondingly, the probability for not being the source is $\frac{M-1}{M}$. 
\item  $\mu$ :  We assume our prior information on the mean of the log-normal distribution is given by $\mbox{NORMAL}(0,10)$. 
\item $\eta$ :  We assume a Gamma distribution with parameters $10$ and $1$ as a prior on $\eta$.
\item $\sigma$ : As noted in the Introduction, we have that $\sigma$ lies in $[18.26,25.72]$. We assume a uniform distribution over the interval. 
\item $M$ : We assume that the value of $M$ is uniform between $100,000$ and $10,000,000$ (see Fig.~\ref{hypergeo}). 
\end{itemize}

\subsubsection{Robustness checks}.

To ensure the robustness of our model to prior selection, we examined a number of alternate priors as well as choices for the prior hyperparameters. For the prior on $\mu$ and $\eta$, we tested a range of alternative hyperparameters that both altered the mean and variance of the prior distributions, finding only slight quantitative shifts in the posterior distribution for the source identification, our target of interest. We also examined the model assumption of a log-normal distribution for flux distributions, testing the model also with a log-Gamma distribution. Again, we found only very slight changes to the posterior distribution of the source identification. 

\subsection{Inference}
Since the posterior cannot be found analytically, we turn to numerical approximations to infer the model.
We use a standard set of Markov chain Monte Carlo (MCMC) routines to sample from the posterior distribution.
Three of the parameters ($\mu$, $\eta$, and $\sigma$) are uncertainty parameters and so are ultimately integrated out to resolve the posterior marginal distribution for $\boldsymbol{z}$, our quantity of interest.
In describing the Metropolis-Hastings (MH) moves, we label the proposed parameter with an asterisk and the current parameter state with no annotation, i.e., $(\lambda^*, \lambda)$.
Rather than write out the full observational model, we use $\pi(D_R,D_S| \mu,\eta, \sigma,M)$ to summarize Eq.~\ref{likelihood}.

The parameter $M$ is qualitatively distinct from the other parameters and numerical investigations indicate that there is strong posterior certainty that the value falls close to the maximum likelihood value. 
To increase the rate of convergence of the MCMC procedure, we approximate the integration over the probability distribution for $M$ as a point mass at $M=1,657,244$, the maximum likelihood value. This approximation vastly speeds up the inference procedure with no apparent change in the inferential quality of the remaining parameters. 

\subsubsection{MCMC updates}
\noindent $\boldsymbol{\sigma}$ : We employ a MH move that proposes a new $\sigma$ value from a uniform distribution with boundaries $18.26$ and $25.72$, the range of values consistent with the specifications of the neutrino detector.
Since the density of choosing a particular value is the same,  the proposal ratio in the MH ratio becomes the ratio of the likelihood.
As $\sigma$ affects the likelihood only through the Rayleigh distribution on a single term, the MH further reduces to the ratio to the ratio of the Rayleigh distributions at the radius of $r_k$ for $z_k=1$: $\alpha = \frac{\mbox{RAYLEIGH}(r_k | \sigma^*) }{\mbox{RAYLEIGH}(r_k | \sigma)}$.\\

\noindent $\boldsymbol{\mu}$ : We propose new values of $\mu$ according to the prior distribution on $\mu$, $\mbox{NORMAL}(0,10)$.
Drawing from the prior simplifies the MH ratio:
\begin{eqnarray*} \alpha & = & \frac{ \pi (D_R,D_S| \mu^*,\eta,\sigma) \cdot \mbox{NORMAL}(\mu|0,10) }{  \pi (D_R,D_S| \mu,\eta,\sigma) \cdot \mbox{NORMAL}(\mu^*|0,10)} \mbox{.}
\end{eqnarray*} 

\noindent $\boldsymbol{\eta}$ :  For $\eta$, we also draw from the prior $\mbox{GAMMA}(10,1)$. By an identical simplification as with $\\mu$, the MH ratio becomes
$$ \alpha  = \frac{ \pi(D_R,D_S|\mu,\eta^*,\sigma) \cdot \mbox{GAMMA}(\eta| 10,1)}{ \pi(D_R,D_S|\mu,\eta,\sigma) \cdot \mbox{GAMMA}(\eta^*| 10,1)} \mbox{.}$$

\noindent $\boldsymbol{z}$ : To draw samples from $\boldsymbol{z}$, we employ a Gibbs update as the posterior provides a convenient conditional formulation: 

\begin{equation}
\begin{split}
 \pi( z_i=1 | D_S,D_R,  \mu, \eta, \sigma )\\ \propto \pi (D_S,D_R | z_i=1, \mu,\eta,\sigma ) \cdot  \pi ( \boldsymbol{z} )  \\
  =  \frac{ \pi (D_S,D_R | z_i=1, \mu,\eta,\sigma ) \cdot   \pi ( \boldsymbol{z} ) }{\sum_{k=1}^K \pi (D_S,D_R | z_i=1, \mu,\eta,\sigma ) \cdot   \pi ( \boldsymbol{z} ) }  \\
\end{split}
\end{equation}
where $\pi(\boldsymbol{z}) = \frac{1}{M}$ if $i \in S \cup R$ or $\frac{M-|R|-|S|}{M}$ if $i \notin S \cup R$. 

\subsubsection{Implementation}
We implement the MCMC in \verb+R+ code.  We find that the resulting algorithm executes 50,000 iterations in approximately 30 minutes, with the effective sample sizes greater than 1000.
Using 10\% burn-in and comparing multiple iterations, the MCMC procedure recovers nearly identical distributions for each of the parameters indicating rapid convergence. 

\subsection{Results}
Drawing $50,000$ samples from the posterior distribution, we estimate the following probabilities for each of point source being the neutrino source. We discard the first $10,000$ iterations as burn-in.
For the remaining, each iteration $t=1,\cdots,40,000$ in the MCMC, we have a corresponding sample $\boldsymbol{z}^{(t)}$. We can then estimate the posterior probability of the source as:
$$ \mathbb{P}(\mbox{source is } k) = \frac{1}{T} \sum_{t=1}^N \boldsymbol{1}_{z_k^{(t)}=1 } \mbox{,}$$
where $\boldsymbol{1}_{A}$ is an indicator random variable if the event $A$ occurs.
The posterior probabilities for each point source are given in Table~\ref{tab:res-stat}. The results indicate some positive probability of support for all of the objects within the \Swift/XRT observation window, as well as positive support for three nearby sources not within the \Swift/XRT field of view.

To garner a definitive estimate of the weight the data brings to bear on the hypothesis that any particular source is the neutrino source, we calculate the Bayes factor. 
Define $H_k$ to be the hypothesis that the neutrino source is at location $k$ and $H_{-k}$ as the hypothesis that the source is not the point source $k$. Then, the Bayes factor can be extracted from the following relationship:

$$ \frac{ \mathbb{P}(H_k|D) }{\mathbb{P}(H_{-k}|D)} = BF_{k,-k}  \frac{\mathbb{P}(H_k)}{\mathbb{P}(H_{-k})}~~ \mbox{,}$$

\noindent where $BF_{k,-k}$ is the Bayes factor between the hypothesis $H_k$ and the hypothesis $H_{-k}$.  The prior probability of $H_k$ is $\frac{1}{K}$ while $H_{-k}$ is $\frac{K-1}{K}$ for any $K$ of the candidate objects. As the undetected objects cannot be distinguished in the model, the prior probability for each of them is $\frac{M-|R|-|S|}{M}$ to indicate that the source is one of these objects. 
The posterior probability $\mathbb{P}(H_k|D)$ is the probability that $z_k=1$, while $\mathbb{P}(H_{-k})$ is the probability that $z_k \neq 0$.
Both of these can be calculated directly using the MCMC samples, as above.
The resulting Bayes factors are listed in Table~\ref{tab:res-stat}. According to the standard interpretation of Bayes factor, any value over 30 makes for very strong evidence for a hypothesis so we can conclude that the data overwhelming supports Source A as the most likely counterpart to the neutrino emission event.

\begin{table}[!ht]
\begin{center}
\caption{Posterior probabilities and Bayes factors of being the neutrino source for each point location according to the MCMC.}
\label{tab:res-stat}
\begin{tabular}{ccc}
Source & Probability & Bayes factor\\
\hline A & 0.89745 & 87.515\\
B  &  0.03399 & 0.345\\
D & 0.02324 & 0.28\\
i & 0.03717 & 0.386\\
ii &  0.00567 & 0.057\\
iv & 0.00285 & 0.028\\
v  & 0.00057 & 0.001\\
vi & 0.00025 & 0.001$<$\\
vii & 0.00015 & 0.001$<$\\
J093305.6+035648 & 0.0002 & $0.0001<$\\
J093149.8+034632 & 0.0005 & $0.0001<$\\
J093117.5+033146 & 0.00008 & $0.0001<$\\
\end{tabular}
\end{center}
\end{table}

\section{Conclusion}
\label{sec:concl}
We have investigated the IceCube neutrino event IceCube-200615A, by performing multiwavelength follow-up and conducting a Bayesian statistical analysis. Our \Swift/XRT tiling observations identified 9 X-ray sources, 5 of which are spurious detections. Out of the 4 remaining bright sources, 1 is an X-ray emitting star, which cannot explain the neutrino event. The three remaining sources are all active galaxies, likely radio-quiet.
An additional 6 sources are detected by ROSAT and are investigated as possible counterparts.
We identified the X-ray brightest counterpart, \rxs, as the likely neutrino emitter. \rxs is a radio-quiet AGN, though it is unclear whether it is jetted with a weak jet like the neutrino emitter NGC\,1068.
In addition to Ice Cube-190331A \citep{Krauss2020} and NGC\,1068 \citep{ngc1068}, this is the third high-energy event with a promising association between an IceCube neutrino event and a radio quiet AGN.
Expected maximum neutrino numbers for \rxs are in the range of $N_{\nu,\mathrm{max}} = 0.066$ -- $0.471$, depending on the SED model and the neutrino spectral index. This is an overestimation due to the assumption of a purely hadronic origin of the high-energy peak of the SED. Despite the relatively low numbers, we note that other possible counterparts are at lower fluxes, and therefore at lower possible neutrino fluxes.
Assuming the neutrino event is physical and is among X-ray detectable sources, the probability that it was produced by source A is 87.5\%, a much higher probability than any other sources.
The other AGN are also possible counterparts to IceCube-200615A, but we note that they are active galaxies, likely radio-quiet, further strengthening the associations between IceCube neutrino events and radio-quiet AGN.

\section*{Acknowledgements}
F.M., N.S., and W.H. were supported by NNH20ZDA001N-NUSTAR through grant \#80NSSC22K1423. G.T., Y.L., and A.P were supported by NNH19ZDA001N-NUSTAR through grants \#80NSSC21K0056 and \#80NSSC21K0057. T.D.R. is an IAF fellow. 
This research has made use of a collection of ISIS functions (ISISscripts) provided by ECAP/Remeis observatory and MIT (http://www.sternwarte.uni-erlangen.de/isis/).

\section*{Data Availability}
\nus, \Swift, \Fermi, VLA data and the ROSAT survey are publicly available.

\bibliographystyle{jwaabib}
\bibliography{abbrv,mnemonic,all}

\begin{thebibliography}{}

\bibitem[\protect\astroncite{{Aartsen} et~al.}{2017a}]{stacking2017}
{Aartsen} M.G., {Abraham} K., {Ackermann} M., et~al., 2017a, ApJ 835, 45

\bibitem[\protect\astroncite{{Aartsen} et~al.}{2017b}]{IC2017}
{Aartsen} M.G., {Ackermann} M., {Adams} J., et~al., 2017b, Journal of
  Instrumentation 12, P03012

\bibitem[\protect\astroncite{{Aartsen} et~al.}{2014}]{Aartsen2014}
{Aartsen} M.G., {Ackermann} M., {Adams} J., et~al., 2014, Phys. Rev. Lett. 113,
  101101

\bibitem[\protect\astroncite{{Abbasi} et~al.}{2024}]{IC-angrecon}
{Abbasi} R., {Ackermann} M., {Adams} J., et~al., 2024, Journal of
  Instrumentation 19, P06026

\bibitem[\protect\astroncite{{Abbasi} et~al.}{2022}]{stacking22}
{Abbasi} R., {Ackermann} M., {Adams} J., et~al., 2022, ApJ 938, 38

\bibitem[\protect\astroncite{{Abbasi} et~al.}{2021}]{IceCube75}
{Abbasi} R., {Ackermann} M., {Adams} J., et~al., 2021, Phys. Rev. D 104, 022002

\bibitem[\protect\astroncite{{Ag{\"u}eros} et~al.}{2009}]{ROSATstars}
{Ag{\"u}eros} M.A., {Anderson} S.F., {Covey} K.R., et~al., 2009, ApJS 181, 444

\bibitem[\protect\astroncite{{Ahn} et~al.}{2012}]{sdss}
{Ahn} C.P., {Alexandroff} R., {Allende Prieto} C., et~al., 2012, ApJS 203, 21

\bibitem[\protect\astroncite{{Albareti} et~al.}{2015}]{Albareti2015}
{Albareti} F.D., {Comparat} J., {Guti{\'e}rrez} C.M., et~al., 2015, Mon. Not.
  R. Astron. Soc. 452, 4153

\bibitem[\protect\astroncite{{Alvarez-Mu{\~n}iz} \&
  {M{\'e}sz{\'a}ros}}{2004}]{AlvarezMuniz2004}
{Alvarez-Mu{\~n}iz} J., {M{\'e}sz{\'a}ros} P.,  2004, Phys. Rev. D 70, 123001

\bibitem[\protect\astroncite{{Atwood} et~al.}{2009}]{fermilat}
{Atwood} W.B., {Abdo} A.A., {Ackermann} M., et~al., 2009, ApJ 697, 1071

\bibitem[\protect\astroncite{{Ayala} \& {HAWC Collaboration}}{2020}]{HAWCAlert}
{Ayala} H., {HAWC Collaboration} 2020, GRB Coordinates Network 27972, 1

\bibitem[\protect\astroncite{{Begelman} et~al.}{1990}]{Begelman1990}
{Begelman} M.C., {Rudak} B., {Sikora} M.,  1990, ApJ 362, 38

\bibitem[\protect\astroncite{{Bell}}{1978}]{Bell1978}
{Bell} A.R.,  1978, Mon. Not. R. Astron. Soc. 182, 147

\bibitem[\protect\astroncite{{Berezinskii} \&
  {Ginzburg}}{1981}]{Berezinskii1981}
{Berezinskii} V.S., {Ginzburg} V.L.,  1981, Mon. Not. R. Astron. Soc. 194, 3

\bibitem[\protect\astroncite{{Beringer} et~al.}{2012}]{Beringer2012}
{Beringer} J., {Arguin} J.F., {Barnett} R.M., et~al., 2012, Phys. Rev. D 86,
  010001

\bibitem[\protect\astroncite{{Bianchi} et~al.}{2011}]{galex}
{Bianchi} L., {Efremova} B., {Herald} J., et~al., 2011, Mon. Not. R. Astron.
  Soc. 411, 2770

\bibitem[\protect\astroncite{{Biermann} \& {Strittmatter}}{1987}]{Biermann1987}
{Biermann} P.L., {Strittmatter} P.A.,  1987, ApJ 322, 643

\bibitem[\protect\astroncite{{Boller} et~al.}{2016}]{2RXS}
{Boller} T., {Freyberg} M.J., {Tr{\"u}mper} J., et~al., 2016, A\&A 588, A103

\bibitem[\protect\astroncite{{B{\"o}ttcher} et~al.}{2013}]{Boettcher2013}
{B{\"o}ttcher} M., {Reimer} A., {Sweeney} K., {Prakash} A.,  2013, ApJ 768, 54

\bibitem[\protect\astroncite{{Brocato} et~al.}{1998}]{nu_stars}
{Brocato} E., {Castellani} V., {degl'Innocenti} S., et~al., 1998, A\&A 333, 910

\bibitem[\protect\astroncite{{CASA Team} et~al.}{2022}]{casanew}
{CASA Team}{Bean} B., {Bhatnagar} S., et~al., 2022, Publ. Astron. Soc. Pac.
  134, 114501

\bibitem[\protect\astroncite{{Chambers} et~al.}{2016}]{panstarrs}
{Chambers} K.C., {Magnier} E.A., {Metcalfe} N., et~al., 2016, arXiv
  arXiv:1612.05560

\bibitem[\protect\astroncite{{Chirkin} \& {Rhode}}{2004}]{Chirkin2004}
{Chirkin} D., {Rhode} W.,  2004, arXiv e-prints  hep--ph/0407075

\bibitem[\protect\astroncite{{Condon} et~al.}{1980}]{Condon1980}
{Condon} J.J., {O'Dell} S.L., {Puschell} J.J., {Stein} W.A.,  1980, Nature 283,
  357

\bibitem[\protect\astroncite{{Cotton} et~al.}{2008}]{Cotton2008}
{Cotton} W.D., {Jaffe} W., {Perrin} G., {Woillez} J.,  2008, A\&A 477, 517

\bibitem[\protect\astroncite{{Dai} \& {Fang}}{2017}]{Dai2017}
{Dai} L., {Fang} K.,  2017, Mon. Not. R. Astron. Soc. 469, 1354

\bibitem[\protect\astroncite{{Dermer} \& {Schlickeiser}}{1993}]{Dermer1993}
{Dermer} C.D., {Schlickeiser} R.,  1993, ApJ 416, 458

\bibitem[\protect\astroncite{{DeYoung} \& {HAWC Collaboration}}{2012}]{hawc}
{DeYoung} T., {HAWC Collaboration} 2012, Nuclear Instruments and Methods in
  Physics Research A 692, 72

\bibitem[\protect\astroncite{{Di Matteo}}{1998}]{diMatteo1998}
{Di Matteo} T.,  1998, Mon. Not. R. Astron. Soc. 299, L15

\bibitem[\protect\astroncite{{Eichler}}{1979}]{Eichler1979}
{Eichler} D.,  1979, ApJ 232, 106

\bibitem[\protect\astroncite{{Evans} et~al.}{2020}]{AMONAlert}
{Evans} P.A., {Gregoire} T., {Kennea} J.A., et~al., 2020, GRB Coordinates
  Network 27973, 1

\bibitem[\protect\astroncite{{Farrar} \& {Gruzinov}}{2009}]{Farrar2009}
{Farrar} G.R., {Gruzinov} A.,  2009, ApJ 693

\bibitem[\protect\astroncite{{Gaia Collaboration} et~al.}{2021}]{gaia}
{Gaia Collaboration}{Brown} A.G.A., {Vallenari} A., et~al., 2021, A\&A 649, A1

\bibitem[\protect\astroncite{{Gao} et~al.}{2019}]{Gao2019}
{Gao} S., {Fedynitch} A., {Winter} W., {Pohl} M.,  2019, Nature Astronomy 3, 88

\bibitem[\protect\astroncite{{Gao} et~al.}{2017}]{Gao2017}
{Gao} S., {Pohl} M., {Winter} W.,  2017, ApJ 843, 109

\bibitem[\protect\astroncite{{Garrappa} et~al.}{2020}]{FermiAlert}
{Garrappa} S., {Buson} S., {Fermi-LAT Collaboration} 2020, GRB Coordinates
  Network 27970, 1

\bibitem[\protect\astroncite{{Gehrels} et~al.}{2004}]{swift}
{Gehrels} N., {Chincarini} G., {Giommi} P., et~al., 2004, ApJ 611, 1005

\bibitem[\protect\astroncite{{Ginzburg} \& {Syrovatskii}}{1965}]{Ginzburg1965}
{Ginzburg} V.L., {Syrovatskii} S.I.,  1965, ARA\&A 3, 297

\bibitem[\protect\astroncite{{Halzen} \& {Hooper}}{2002}]{Halzen2002}
{Halzen} F., {Hooper} D.,  2002, Reports on Progress in Physics 65, 1025

\bibitem[\protect\astroncite{{Halzen} \& {Hooper}}{2005}]{Halzen2005}
{Halzen} F., {Hooper} D.,  2005, Astroparticle Physics 23, 537

\bibitem[\protect\astroncite{{Harrison} et~al.}{2013}]{nustar}
{Harrison} F.A., {Craig} W.W., {Christensen} F.E., et~al., 2013, ApJ 770, 103

\bibitem[\protect\astroncite{{HI4PI Collaboration} et~al.}{2016}]{HI4PI}
{HI4PI Collaboration}{Ben Bekhti} N., {Fl{\"o}er} L., et~al., 2016, A\&A 594,
  A116

\bibitem[\protect\astroncite{{Houck} \& {Denicola}}{2000}]{ISIS}
{Houck} J.C., {Denicola} L.A.,  2000,
\newblock In: {Manset} N., {Veillet} C., {Crabtree} D. (eds.) Astronomical Data
  Analysis Software and Systems IX, 216. Astronomical Society of the Pacific
  Conference Series, p. 591

\bibitem[\protect\astroncite{{IceCube Collaboration}}{2013}]{IC2013}
{IceCube Collaboration} 2013, Science 342, 1242856

\bibitem[\protect\astroncite{{IceCube Collaboration}}{2018}]{IC2018}
{IceCube Collaboration} 2018, Science 361, 147

\bibitem[\protect\astroncite{{IceCube Collaboration}}{2020}]{IceCubeAlert}
{IceCube Collaboration} 2020, GRB Coordinates Network 27950, 1

\bibitem[\protect\astroncite{{IceCube Collaboration}}{2022}]{ngc1068}
{IceCube Collaboration} 2022, Science 378, 538

\bibitem[\protect\astroncite{{IceCube Collaboration}}{2023}]{ICgal}
{IceCube Collaboration} 2023, Science 380, 1338

\bibitem[\protect\astroncite{{IceCube Collaboration} et~al.}{2006}]{ic1}
{IceCube Collaboration}{Achterberg} A., {Ackermann} M., et~al., 2006,
  Astroparticle Physics 26, 155

\bibitem[\protect\astroncite{{IceCube Collaboration} et~al.}{2018}]{txs0506}
{IceCube Collaboration}{Fermi-LAT Collaboration, and }{MAGIC
  Collaboration}et~al., 2018, Science 361, eaat1378

\bibitem[\protect\astroncite{{Kadler} et~al.}{2016}]{Kadler2016}
{Kadler} M., {Krau{\ss}} F., {Mannheim} K., et~al., 2016, Nature Physics 12,
  807

\bibitem[\protect\astroncite{{Katgert} et~al.}{1973}]{Katgert1973}
{Katgert} P., {Katgert-Merkelijn} J.K., {Le Poole} R.S., {van der Laan} H.,
  1973, A\&A 23, 171

\bibitem[\protect\astroncite{{Keivani} et~al.}{2018}]{Keivani2018}
{Keivani} A., {Murase} K., {Petropoulou} M., et~al., 2018, ApJ 864, 84

\bibitem[\protect\astroncite{{Kellermann} et~al.}{1989}]{Kellerman1989}
{Kellermann} K.I., {Sramek} R., {Schmidt} M., et~al., 1989, AJ 98, 1195

\bibitem[\protect\astroncite{{Kheirandish} et~al.}{2021}]{Kheirandish2021}
{Kheirandish} A., {Murase} K., {Kimura} S.S.,  2021, ApJ 922, 45

\bibitem[\protect\astroncite{{Krau{\ss}} et~al.}{2020a}]{ic1903}
{Krau{\ss}} F., {Calamari} E., {Keivani} A., et~al., 2020a, Mon. Not. R.
  Astron. Soc. 497, 2553

\bibitem[\protect\astroncite{{Krau{\ss}} et~al.}{2020b}]{Krauss2020}
{Krau{\ss}} F., {Calamari} E., {Keivani} A., et~al., 2020b, Mon. Not. R.
  Astron. Soc. 497, 2553

\bibitem[\protect\astroncite{{Krau{\ss}} et~al.}{2018}]{Krauss2018}
{Krau{\ss}} F., {Deoskar} K., {Baxter} C., et~al., 2018, A\&A 620, A174

\bibitem[\protect\astroncite{{Krau{\ss}} et~al.}{2014}]{Krauss2014}
{Krau{\ss}} F., {Kadler} M., {Mannheim} K., et~al., 2014, A\&A 566, L7

\bibitem[\protect\astroncite{{Krau{\ss}} et~al.}{2016}]{Krauss2016}
{Krau{\ss}} F., {Wilms} J., {Kadler} M., et~al., 2016, A\&A 591, A130

\bibitem[\protect\astroncite{{Loeb} \& {Waxman}}{2006}]{Loeb2006}
{Loeb} A., {Waxman} E.,  2006, JCAP 2006, 003

\bibitem[\protect\astroncite{{Magnier} et~al.}{2020}]{panstarrs2}
{Magnier} E.A., {Chambers} K.C., {Flewelling} H.A., et~al., 2020, ApJS 251, 3

\bibitem[\protect\astroncite{{Mannheim}}{1993}]{Mannheim1993}
{Mannheim} K.,  1993, A\&A 269, 67

\bibitem[\protect\astroncite{{Mannheim}}{1995}]{Mannheim1995}
{Mannheim} K.,  1995, Astroparticle Physics 3, 295

\bibitem[\protect\astroncite{{Mannheim} \& {Biermann}}{1992}]{Mannheim1992}
{Mannheim} K., {Biermann} P.L.,  1992, A\&A 253, L21

\bibitem[\protect\astroncite{{Maraschi} et~al.}{1992}]{Maraschi1992}
{Maraschi} L., {Ghisellini} G., {Celotti} A.,  1992, ApJL 397, L5

\bibitem[\protect\astroncite{{Marshall} et~al.}{2020}]{UVOTAlert}
{Marshall} F.E., {Evans} P.A., {Gregoire} T., et~al., 2020, GRB Coordinates
  Network 27979, 1

\bibitem[\protect\astroncite{{Massaro} et~al.}{2004}]{Massaro2004}
{Massaro} E., {Perri} M., {Giommi} P., {Nesci} R.,  2004, A\&A 413, 489

\bibitem[\protect\astroncite{{Merloni} et~al.}{2024}]{erass}
{Merloni} A., {Lamer} G., {Liu} T., et~al., 2024, A\&A 682, A34

\bibitem[\protect\astroncite{{M{\"u}cke} et~al.}{2000}]{Muecke2000}
{M{\"u}cke} A., {Engel} R., {Rachen} J.P., et~al., 2000, Computer Physics
  Communications 124, 290

\bibitem[\protect\astroncite{{Murase} et~al.}{2013}]{Murase2013}
{Murase} K., {Ahlers} M., {Lacki} B.C.,  2013, Phys. Rev. D 88, 121301

\bibitem[\protect\astroncite{{Murase} et~al.}{2016}]{Murase2016}
{Murase} K., {Guetta} D., {Ahlers} M.,  2016, Phys. Rev. Lett. 116, 071101

\bibitem[\protect\astroncite{{Murase} \& {Ioka}}{2013}]{Murase2013b}
{Murase} K., {Ioka} K.,  2013, Phys. Rev. Lett. 111, 121102

\bibitem[\protect\astroncite{{Murase} et~al.}{2020}]{Murase2020}
{Murase} K., {Kimura} S.S., {M{\'e}sz{\'a}ros} P.,  2020, Phys. Rev. Lett. 125,
  011101

\bibitem[\protect\astroncite{{Nasa High Energy Astrophysics Science Archive
  Research Center (Heasarc)}}{2014}]{heasoft}
{Nasa High Energy Astrophysics Science Archive Research Center (Heasarc)} 2014,
\newblock {HEAsoft: Unified Release of FTOOLS and XANADU},
\newblock Astrophysics Source Code Library, record ascl:1408.004

\bibitem[\protect\astroncite{{Norman} et~al.}{1995}]{Norman1995}
{Norman} C.A., {Melrose} D.B., {Achterberg} A.,  1995, ApJ 454, 60

\bibitem[\protect\astroncite{{Nowak} et~al.}{2012}]{Nowak2012}
{Nowak} M.A., {Neilsen} J., {Markoff} S.B., et~al., 2012, ApJ 759, 95

\bibitem[\protect\astroncite{{Protheroe} \& {Kazanas}}{1983}]{Protheroe1983}
{Protheroe} R.J., {Kazanas} D.,  1983, ApJ 265, 620

\bibitem[\protect\astroncite{{Rachen} et~al.}{1993}]{Rachen1993}
{Rachen} J.P., {Stanev} T., {Biermann} P.L.,  1993, A\&A 273, 377

\bibitem[\protect\astroncite{{Rees}}{1967}]{Rees1967}
{Rees} M.J.,  1967, Mon. Not. R. Astron. Soc. 137, 429

\bibitem[\protect\astroncite{{Roming} et~al.}{2005}]{uvot}
{Roming} P.W.A., {Kennedy} T.E., {Mason} K.O., et~al., 2005, Space Sci. Rev.
  120, 95

\bibitem[\protect\astroncite{{Sikora} et~al.}{1994}]{Sikora1994}
{Sikora} M., {Begelman} M.C., {Rees} M.J.,  1994, ApJ 421, 153

\bibitem[\protect\astroncite{{Skrutskie} et~al.}{2006}]{2MASS}
{Skrutskie} M.F., {Cutri} R.M., {Stiening} R., et~al., 2006, AJ 131, 1163

\bibitem[\protect\astroncite{{Smith} et~al.}{2013}]{amon}
{Smith} M.W.E., {Fox} D.B., {Cowen} D.F., et~al., 2013, Astroparticle Physics
  45, 56

\bibitem[\protect\astroncite{{Stecker} et~al.}{1991}]{Stecker1991}
{Stecker} F.W., {Done} C., {Salamon} M.H., {Sommers} P.,  1991, Phys. Rev.
  Lett. 66, 2697

\bibitem[\protect\astroncite{{Stein} et~al.}{2021}]{tde}
{Stein} R., {van Velzen} S., {Kowalski} M., et~al., 2021, Nature Astronomy 5,
  510

\bibitem[\protect\astroncite{{Stocke} et~al.}{1992}]{Stocke1992}
{Stocke} J.T., {Morris} S.L., {Weymann} R.J., {Foltz} C.B.,  1992, ApJ 396, 487

\bibitem[\protect\astroncite{{Tavecchio} et~al.}{2000}]{Tavecchio2000}
{Tavecchio} F., {Maraschi} L., {Sambruna} R.M., {Urry} C.M.,  2000, ApJL 544,
  L23

\bibitem[\protect\astroncite{{Thompson} et~al.}{1980}]{vla}
{Thompson} A.R., {Clark} B.G., {Wade} C.M., {Napier} P.J.,  1980, ApJS 44, 151

\bibitem[\protect\astroncite{{Verner} et~al.}{1996}]{Verner1996}
{Verner} D.A., {Ferland} G.J., {Korista} K.T., {Yakovlev} D.G.,  1996, ApJ 465,
  487

\bibitem[\protect\astroncite{{Vietri}}{1998}]{Vietri1998}
{Vietri} M.,  1998, ApJ 507, 40

\bibitem[\protect\astroncite{{Waters} et~al.}{2020}]{panstarrs3}
{Waters} C.Z., {Magnier} E.A., {Price} P.A., et~al., 2020, ApJS 251, 4

\bibitem[\protect\astroncite{{Waxman} \& {Bahcall}}{1997}]{Waxman1997}
{Waxman} E., {Bahcall} J.,  1997, Phys. Rev. Lett. 78, 2292

\bibitem[\protect\astroncite{{Wilms} et~al.}{2000}]{Wilms2000}
{Wilms} J., {Allen} A., {McCray} R.,  2000, ApJ 542, 914

\bibitem[\protect\astroncite{{Wright} et~al.}{2010}]{wise}
{Wright} E.L., {Eisenhardt} P.R.M., {Mainzer} A.K., et~al., 2010, AJ 140, 1868

\end{thebibliography}

\bsp	
\label{lastpage}
\end{document}